\documentclass{biophys-new}
\usepackage[utf8]{inputenc}
\usepackage{graphicx}
\usepackage{units}
\usepackage[colorlinks,allcolors=cyan!70!black]{hyperref}

\usepackage{lipsum}

\title{Ion Condensation onto Ribozyme is Site-Specific and Fold-Dependent}
\runningtitle{Ion Condensation onto Ribozyme} 

\author[1]{Naoto Hori}
\author[2]{Natalia A. Denesyuk}
\author[1,*]{D. Thirumalai}
\runningauthor{Hori, Denesyuk, Thirumalai} 

\affil[1]{Department of Chemistry, University of Texas, Austin, Texas 78712, USA}
\affil[2]{Biophysics Program, Institute for Physical Science and Technology, University of Maryland, College Park, Maryland 20742, USA}

\corrauthor[*]{dave.thirumalai@gmail.com}

\papertype{Article}

\newcommand{\Mg}{Mg\textsuperscript{2+}}
\newcommand{\Ca}{Ca\textsuperscript{2+}}
\newcommand{\K}{K\textsuperscript{+}}
\newcommand{\Cl}{Cl\textsuperscript{-}}

\begin{document}

\begin{frontmatter}

\begin{abstract}
The highly charged RNA molecules, with each phosphate carrying a single negative charge, cannot fold into well-defined architectures with tertiary interactions, in the absence of ions. For ribozymes, divalent cations are known to be more efficient than monovalent ions in driving them to a compact state although \Mg{} ions are needed for catalytic activities. Therefore, how ions interact with RNA is relevant in understanding RNA folding. It is often thought that most of the ions are territorially and non-specifically bound to the RNA, as predicted by the counterion condensation (CIC) theory. Here, we show using simulations of {\it Azoarcus} ribozyme, based on an accurate coarse-grained Three Site Interaction (TIS) model, with explicit divalent and monovalent cations, that ion condensation is highly specific and depends on the nucleotide position. The regions with high coordination between the phosphate groups and the divalent cations are discernible even at very low \Mg{} concentrations when the ribozyme does not form tertiary interactions. Surprisingly, these regions also contain the secondary structural elements that nucleate subsequently in the self-assembly of RNA, implying that ion condensation is determined by the architecture of the folded state. These results are in sharp contrast to interactions of ions (monovalent and divalent) with rigid charged rods in which ion condensation is uniform and position independent. The differences are explained in terms of the dramatic non-monotonic shape fluctuations in the ribozyme as it folds with increasing \Mg{} or \Ca{} concentration.
\end{abstract}

\end{frontmatter}

\section*{Introduction}
RNA molecules are negatively charged polyelectrolytes (PEs) with each molecule carrying a total charge equal to $-Ne$ where $N$ is the number of nucleotides and $e$ is the elementary charge. 
The negative charges are localized on the phosphate groups. Because of the large electrostatic repulsion between the phosphate groups, RNA can only fold if the effective charge on the phosphate group is reduced or renormalized \cite{Sun17ARB,Tao18BJ,Pollack11Biopolymers,Kirmizialtin12BJ,Denesyuk13JPCB}, which we know  occurs by the counterion condensation (CIC) mechanism \cite{Oosawa57JPolymerSci,Manning69JCP1,Oosawa1971Book,Alexander84JCP}.
In this process, described decades ago in remarkable studies by Oosawa and Manning \cite{Oosawa57JPolymerSci,Manning69JCP1}, the condensed ions are localized in an apparent volume occupied by the PE or any highly charged macroion. 
As a consequence, the effective charge on the negatively charged moieties in the PE (or the phosphate groups in DNA and RNA) is reduced from $-e$ to $-\beta e$ with $\beta<1$ \cite{Oosawa57JPolymerSci,Manning69JCP1,Oosawa1971Book}.
The renormalized charge parameter, $\beta$, can be calculated for PEs with regular shapes \cite{Oosawa1971Book}.
For example, the condition for ion condensation for a rod-like macroion is determined by the Manning parameter $\xi=\ell_{\mathrm{B}}/b>\nicefrac{1}{\lvert z\rvert}$, where $\ell_{\mathrm{B}}$ is the Bjerrum length, $b$ is the mean distance between charges on the polyelectrolyte, and $z$ is the valence of the counterion.
The Bjerrum length is the distance at which the electrostatic energy of two univalent charges is equal to the thermal energy, $\ell_{\mathrm{B}}=e^{2}\left(4\pi\varepsilon_{0}\varepsilon k_{\textrm{B}}T\right)^{-1}$, where $\varepsilon_{0}$ is the vacuum permittivity, $\varepsilon$ is the dielectric constant of the solvent, and $k_{\textrm{B}}T$ is the thermal energy.
For highly charged rods, ion condensation occurs when $\xi \ge \nicefrac{1}{\lvert z\rvert}$, with the transition being sharp for long thin rods (high aspect ratio). 
A similar condition may be obtained for spherical macroions. In this case, ion condensation does occur with a fraction localized in the vicinity of the sphere, but unlike the situation of the cylindrical macroion, there is no critical value of $\xi$ at which counterion condensation is initiated \cite{Oosawa1971Book}.

There are two implications of these well-known aspects of the CIC theory, summarized above, that are relevant to the current study.
(1) Monovalent ions ($\lvert z\rvert=1$) are not as efficient in charge renormalization as higher-valent ions ($\lvert z\rvert>1$). 
Based on this consideration, we calculated  that about 90\% of the phosphate charge in \textit{Tetrahymena} ribozyme  is neutralized in the presence of divalent and trivalent cations \cite{HeilmanMiller01JMBeq}. 
(2) The condensation mechanism depends critically on the shape of the macroion.  
As noted above, the theoretical predictions are different for spheres and rods. 
Therefore, we expect that the CIC mechanism ought to be different for RNA, which undergoes substantial conformational fluctuations during the ion-driven folding process.
The shape of {\it Azoarcus} ribozyme that is used as an example here, changes as the concentration of ions (\Mg{}) is increased from a low to a high value. 
This ribozyme, and many other RNA molecules, are not spherical even when folded \cite{Hyeon06JCP}. Moreover,
even at low \Mg{} concentrations, {\it Azoarcus} ribozyme does not adopt globally cylindrical structures although individual helices may be rigid enough to be pictured as stiff small cylinders. 

Because of the irregular shapes of RNA molecules at all ion concentrations, ion-RNA interactions, which are often thought to involve diffuse ions that are territorially (within a volume of RNA where the electrostatic potential is substantial) but non-specifically bound, require scrutiny.
Motivated by the considerations described above, we answer the following question here: How
does CIC occur in such molecules whose shapes are not only irregular but also change as they fold?
We are able to quantitatively answer this question because of the development of an accurate coarse-grained force field, which has been used to quantitatively predict the thermodynamic properties of {\it Azoarcus} ribozyme and other RNA molecules \cite{Denesyuk15NatChem} in the presence of \K{}, \Mg{} and \Ca{}. 

There is an additional reason for undertaking this investigation.
The physically appealing CIC mechanism qualitatively explains RNA folding both from the stability and kinetic perspective \cite{HeilmanMiller01JMBeq}.
The mechanism is based on the basic assumption that condensed ions are diffuse and freely move within the volume surrounding the RNA where the electrostatic potential is the largest. 
Indeed, the condition for ion condensation is obtained by assuming that the mobile but bound ions are in equilibrium with free ions (the chemical potential of these two species are identical) that explore the region outside the sphere of influence of the macroion \cite{Oosawa1971Book}. 
Such an assumption does qualitatively explain the success of even the $z$-dependent folding kinetics of \textit{Tetrahymena} ribozyme \cite{HeilmanMiller01JMBkin,Thirumalai01AnnRevPhysChem}.
However, our recent studies \cite{Denesyuk15NatChem} suggest that the assumption of uniform condensation, often made in rationalizing electrostatic effects in RNA, may be qualitatively incorrect.
In other words, nucleotide-specific interactions between ions (especially \Ca{} and \Mg{} ions) provide the key in quantitatively describing charge renormalization in ribozyme folding. 

In order to fully understand the nature and consequence of ion condensation onto RNA, we undertook two types of simulations.
We first probed the interaction of divalent cations with small finite-sized rigid rods of different linear charge density, expressed as $b=n/L$ where $n$ is the total number of charges and $L$ is the cylinder length.
It follows that $b$ is the mean distance between charges. However, in our simulations the distance between the charges ($b$) is a constant.
We also investigated the condensation of divalent cations onto {\it Azoarcus} ribozyme using an accurate model introduced recently \cite{Denesyuk15NatChem}.
By comparing the results of these two systems, we elaborate on the importance of the architecture of the macroion (rod versus ribozyme) on the ion condensation mechanism. 
Our major finding is that the shape of the RNA matters for counterion condensation.
In other words, ion-condensation to RNA depends exquisitely on the folded RNA architecture, and hence cannot be treated as a non-specific interaction that is used to understand CIC for macroion with definite simple shapes. A surprise in our study is that the specificity of ion binding plays a vital role in driving RNA folding even at very low divalent ion concentrations.

\begin{figure}[hbt!]
\centering
\includegraphics[height=0.6\textheight]{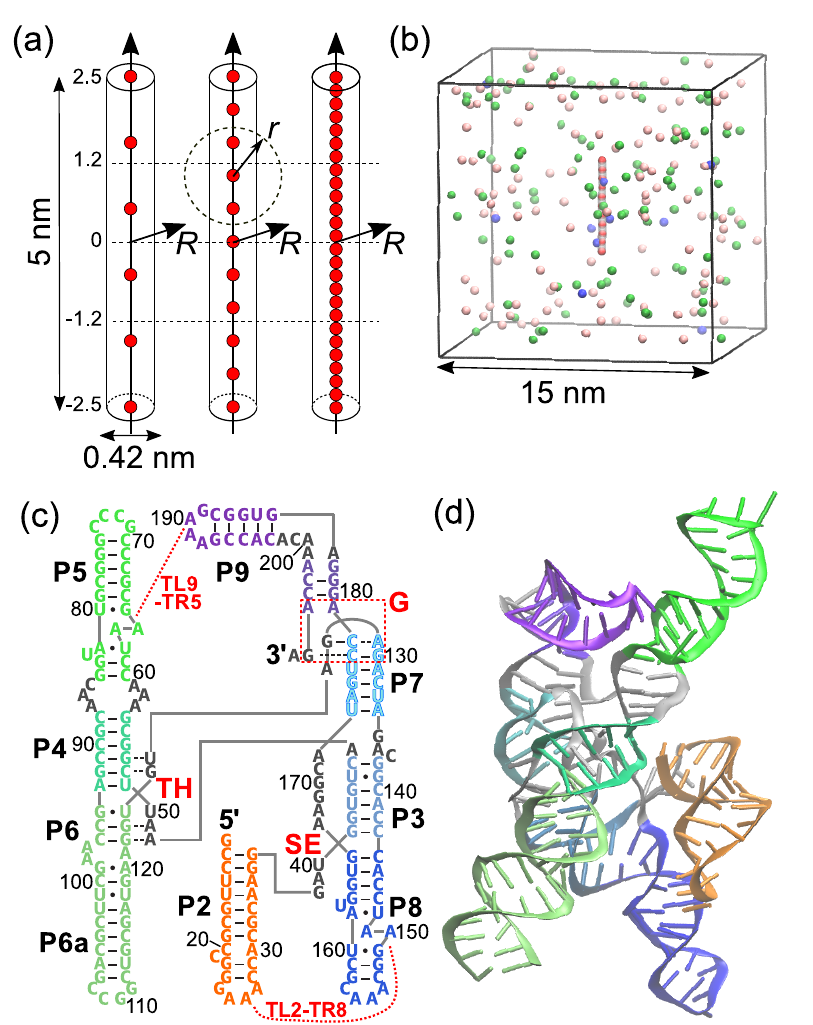}
\caption{
Schematic of charged rods and the {\it Azoarcus} ribozyme.
\textbf{(a)} The dimensions of the rod, with aspect ratio 11.9, and positions of charges (in red) are illustrated. The diameter of the rod is 0.42 nm, which is the same as the diameter of the phosphate group in the coarse-grained RNA model.
Negative charges are placed at regular interval with spacing that gives charge densities; $-1.0$ $e$/nm (left), $-2.0$ $e$/nm (middle), and $-5.0$ $e$/nm (right).
\textbf{$R$} is the radial distance in the cylindrical coordinate, whereas $r$ represents the distance from the point charge to a position in space in the spherical coordinate.
\textbf{(b)} A snapshot of the rod simulations. Ions are colored in blue for \Mg{}, green for \K{}, and pink for \Cl{}.
\textbf{(c, d)} The secondary (c) and tertiary (d) structures of {\it Azoarcus} group I intron ribozyme (PDB 1U6B \cite{AzoXray04N}).
The color codes in both structures are the same.
The locations of the key tertiary interactions are indicated in red: TH, Triple Helix; SE, Stack-Exchange junction; G, G-binding pocket; TL-TR, tetraloop-tetraloop receptor.
}
\label{fig:setup}
\end{figure}

\section*{Model and Methods}

\subsection*{Rigid rod simulations} 
A 5 nm-long rigid cylindrical rod is placed at the center of 15 nm-wide cubic box (Fig.~\ref{fig:setup}a and b).
The diameter of the rod is 0.42 nm, which is the size of  the phosphate bead in the coarse-grained RNA model described below.
Technically, the rod is constructed by overlapping spheres that are separated by 0.1 nm.
The rod is completely rigid and fixed at the center of the simulation box, whereas ions are allowed to move freely around the rod. We use periodic boundary conditions. 
Three different charge densities on the rod, $\rho_{_{\mathrm{1D}}}=$ $-1$, $-2$, and $-5$ $e\,\textrm{nm}^{-1}$ were simulated, which translates to the constant separation distance ($b$) between neighboring charges as 1.0, 0.5, and 0.2 nm, respectively.
For comparison, the Bjerrum length ($\ell_{\mathrm{B}}$) is 0.73 nm at the simulation temperature, $T=37^{\circ}\textrm{C}$.
The interactions between the charges on the rod and the ions are the same as in the coarse-grained RNA-ion model \cite{Denesyuk15NatChem}.
Ions (\K{}, Cl$^{-}$, and \Mg{}) are modeled as spheres and interact with each other and with the charged rod via the excluded volume potential, which is identical to the one used in the RNA model,
\begin{equation}
U_{\textrm{Ex}}=\varepsilon_{ij}\left[\left(\frac{1.6}{r_{ij}+1.6-D_{ij}}\right)^{12}-2\left(\frac{1.6}{r_{ij}+1.6-D_{ij}}\right)^{6}+1\right],\;r_{ij}\leq D_{ij} \label{eq:Uex}
\end{equation}
where $r$ is the distance of the two beads, $D_{ij}=R_{i}+R_{j}$, and $\varepsilon_{ij}=\sqrt{\varepsilon_{i}\varepsilon_{j}}$.
The values of $R_{i}$ and $\varepsilon_{i}$ for the ions and the RNA sites are tabulated in Table S1 in the Supplementary Information (SI). 
The $R_{i}$ and $\varepsilon$ for the charged rod are the same as the phosphate site in the RNA model.
If $r_{ij}>D_{ij}$ then $U_{\textrm{EX}}=0$.
For any pair of charged beads, the standard Coulomb potential is used, 
\begin{equation}
U_{\mathrm{Ele}}=\frac{q_{i}q_{j}}{4\pi\varepsilon_{0}\varepsilon r_{ij}}
\end{equation}
where $q_{i}$ is the charge of the $i$\textsuperscript{th} bead.
The dielectric constant, $\varepsilon$, is temperature dependent and is 74 at the simulation temperature $T=37^{\circ}\mathrm{C}$.
A series of Langevin dynamics simulations were performed for $2\times10^{7}$ iteration time steps, in which ion positions are saved every 100 steps.
Thus, in total $2\times10^{6}$ snapshots are used for analyses for each setup.

We ought to emphasize that the results of rigid rod simulations for different purposes have been reported numerous times in the literature (see for example \cite{GronbechJensen97PRL,Borukhov97PRL,Deserno00Macromolecules}) although new phenomena continue to be found in \cite{Cha18PRL}.
The simulations presented here merely serve as a control in order to illustrate the dramatically different findings for the {\it Azoarcus} ribozyme.

\subsection*{Simulations of group I intron RNA}
Because the model and simulation methods of the ribozyme were described in detail previously \cite{Denesyuk15NatChem}, we give only a brief description here.
We used a coarse-grained Three Interaction Site (TIS) RNA model that has three interaction sites, for each nucleotide, corresponding to phosphate, sugar and base moieties \cite{Hyeon05PNAS}.
All the ions are explicitly treated whereas water is modeled implicitly using the temperature-dependent dielectric constant.
We considered concentration of \Mg{} ions in the range from 0 to 30 mM in the presence of either 50 mM or 12 mM of \K{} that is typically used in the Tris buffer \cite{Behrouzi12Cell}.
We added an appropriate number of anions ($\textrm{Cl}^{-}$) to ensure that the entire system is charge neutral.
All the simulations were conducted using a periodic-boundary cubic box with each side being 35 nm.
We performed a series of Langevin dynamics simulations at 37$^{\circ}\textrm{C}$ in order to efficiently sample conformations of the system containing the ribozyme and ions \cite{Honeycutt92Biopolymers}.
For each salt concentration, several independent simulations were started from the crystal structure and trajectories were generated for sufficiently long time to ensure that equilibrium is reached before we obtained snapshots for analyses.
For low \Mg{}/\Ca{} concentrations, at which the ribozyme is unfolded state, we generated seven trajectories with 7.5 $\mu$s of equilibration followed by 15--22.5 $\mu$s of production runs.
The length of these simulations are long enough to observe multiple unfolding/folding events in the low-friction Langevin dynamics.
For higher \Mg{}/\Ca{} concentrations, at which the ribozyme is predominantly in the folded conformations, we ran three trajectories with length 7.5 $\mu$s and discarded the first $10^5$ steps of each trajectory.
Elsewhere \cite{Denesyuk15NatChem} we provide more detailed descriptions of the model and the simulation protocol.

The crystal structure of \textit{Azoarcus} group I intron (PDB 1U6B~\cite{AzoXray04N}, Fig.~\ref{fig:setup}c and d) was used as the reference structure of the native conformation.
The numbering of nucleotides from G12 through G206 is consistent with the crystal structure ~\cite{AzoXray04N} and the first nucleotide of the exon sequence is labeled 207.

\subsection*{Data analyses}
To quantify ion condensation onto the RNA and the rigid rod, we calculated the \textit{local ion concentration}, a fingerprint of ion interaction with the macroion, around the $i$\textsuperscript{th} phosphate (or point charges on the rod) defined as,
\begin{equation}
c_{i}^{\ast}=\frac{1}{N_{\mathrm{A}}V_{\mathrm{c}}}\int_{0}^{r_{\mathrm{c}}}\rho_{i}(r)\,4\pi r^{2}\mathrm{d}r,\label{eq:localconc}
\end{equation}
where $\rho_{i}(r)$ is the number of ions per unit volume (number
density) of the ion at the distance $r$ from the $i$\textsuperscript{th}
phosphate, $V_{c}$ is the spherical volume of radius $r_{c}$, $N_{\mathrm{A}}$
is the Avogadro's number to represent $c_{i}^{\ast}$ in molar units.
We use the Bjerrum length as the cutoff distance, $r_{\textrm{c}}=\ell_{\mathrm{B}}=0.73\,\textrm{nm}$
because at that distance the favorable Coulomb attraction between
a cation and an anion ($z=1$) is $k_{\mathrm{B}}T$\textit{.} 

We also calculated the fluctuations
in the local ion concentration as the normalized root-mean-square
deviation from the mean, 
\begin{equation}
\Delta c_{i}^{\ast}=\frac{\left\langle \left(c_{i}^{\ast}-\left\langle c_{i}^{\ast}\right\rangle \right)^{2}\right\rangle ^{\nicefrac{1}{2}}}{\left\langle c_{i}^{\ast}\right\rangle },\label{eq:fluctlocalconc}
\end{equation}
where $\left\langle c_{i}^{\ast}\right\rangle $ is the ensemble
average of the local ion concentration around the $i$\textsuperscript{th} nucleotide. 
Thus, $c_{i}^{\ast}-\left\langle c_{i}^{\ast}\right\rangle $ is an instantaneous deviation from the ensemble average.
The average ion concentration is, 
\begin{equation}
\overline{\left\langle c^{\ast}\right\rangle }=\frac{1}{N}\sum_{i=1}^{N}\left\langle c_{i}^{\ast}\right\rangle ,\label{eq:averagelocalconc}
\end{equation}
where $N$ is the number of nucleotides. 
Note that the upper bar indicates an average over the sequence, whereas the bracket indicates the ensemble average.
The extent of deviation of ion concentration fluctuation from $\left\langle c^{\ast}\right\rangle $ along the sequence was calculated using, 
\begin{equation}
\sigma^{2}=\frac{1}{N}\sum_{i=1}^{N}\left(\left\langle c_{i}^{\ast}\right\rangle -\overline{\left\langle c^{\ast}\right\rangle }\right)^{2}.\label{eq:deviationlocalconc}
\end{equation}
In the results section, for simplicity, we omit the brackets for the ensemble average. 
Thus, all quantities are averaged over an ensemble of configurations in the system (ions and the charged RNA or the rod) unless otherwise stated.

For charged rods, we also computed the total charge per unit length due to contributions from each ionic species.
The effective charge densities around the rod were calculated as the amount of accumulated charge per unit length for the ion species $m$,
\begin{equation}
Q_{m}(R)=z_{m}\int_{0}^{R}\rho_{m}(R')2\pi R'\,\mathrm{d}R',\label{eq:Q}
\end{equation}
where $\rho_{m}(R)$ is the number density of species $m$ at the radial distance $R$ from the axis of the cylinder, and $z_{m}$ is its valence. 
Note that $Q$ can be  computed for each ion \Mg{}, \K{}, and Cl$^{-}$ independently. Finally, the total effective charge per unit length is the sum of the rod charge and contributions from those ions,
\begin{equation}
Q_{\mathrm{T}}(R)=\rho_{_{\mathrm{1D}}}+Q_{\mathrm{Mg^{2+}}}(R)+Q_{\mathrm{K^{+}}}(R)+Q_{\mathrm{Cl^{-}}}(R).\label{eq:Qtotal}
\end{equation}
Note that we use $R$ for the radial distance from the center of the rods in the cylindrical coordinate, whereas $r$ is for the distance from the point charges on the rod in the spherical coordinate (see Fig.~\ref{fig:setup}a). 

\section*{Results}

\subsection*{Charged cylindrical rods}
\begin{figure}[hbt!]
\centering
\includegraphics[width=0.6\textwidth]{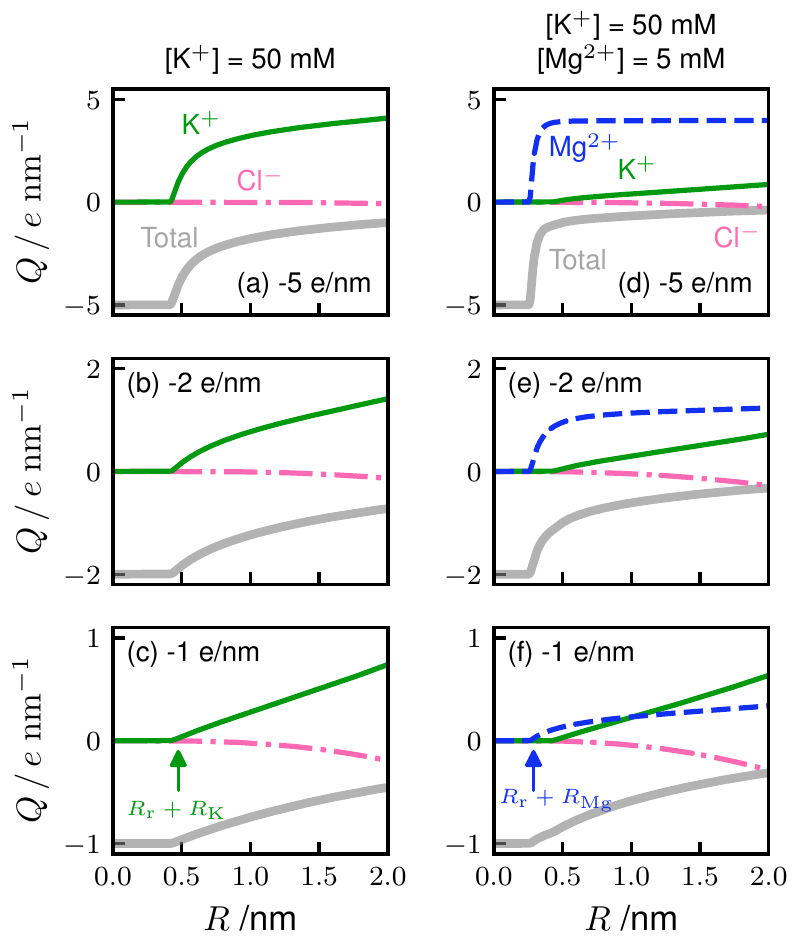}
\caption{ 
Comparison of the condensed ion species on charged rods with linear densities, $\rho_{_{\mathrm{1D}}}=$ $-5\,e/\mathrm{nm}$ (top panels), $-2\,e/\mathrm{nm}$ (middle) and $-1\,e/\mathrm{nm}$ (bottom) in the absence (a-c) and presence (d-f) of \Mg{}. 
The cumulative charges per unit length $Q$ (Eq.~\ref{eq:Q}) is plotted as a function of $R$ for contribution from each ionic species (annotated in the top panels).
The total cumulative charge, $Q_{\mathrm{T}}(R)$ (thick gray line), is the sum of charges of the rod and surrounding ions (Eq.~\ref{eq:Qtotal}). 
The solution conditions are indicated on the top.
The ranges of exclusion by the volume interactions (Eq.~\ref{eq:Uex}), $R_{\mathrm{r}} + R_{\mathrm{K}} = 0.48$~nm for \K{}, and $R_{\mathrm{r}}+R_{\mathrm{Mg}}=0.29$~nm for  \Mg{}, are indicated by arrows in the bottom panels ($R_{\mathrm{r}}$, $R_{\mathrm{K}}$, $R_{\mathrm{Mg}}$ are radii of the rod, \K{}, and \Mg{}, respectively). 
\label{fig:rod_radial}
}
\end{figure}

We first describe condensations of ions onto charged rods with varying spacing between fixed charges, $b=-e/\rho_{_{\mathrm{1D}}}$, where $\rho_{_{\mathrm{1D}}}$ is the charge density of the rod. 
Based on the CIC theory, we expect that ions should condense uniformly along the rod.
The condition for condensation in the CIC theory is that $\xi>\nicefrac{1}{\lvert z\rvert}$, where $z$ is the valence of the ion species. In the absence of multivalent salt, $\xi>1$ and thus we expect ions should condense on rods when $\rho_{_{\mathrm{1D}}}=$ $-5\,e/\mathrm{nm}$ ($\xi=3.6$) and $\rho_{_{\mathrm{1D}}}=$ $-2\,e/\mathrm{nm}$ ($\xi=1.5$) among the three values of charge densities we examine here.
This expectation was borne out in the simulations even though the aspect ratio ($\approx$ 11.9) of the rod was relatively small.
In Fig.~\ref{fig:rod_radial}(a-c), the cumulative charges, $Q$ (Eq.~\ref{eq:Q}), around the rod are plotted for each ion species and the net charge.
The condensation of monovalent ions appears as a sharp increase in $Q_{\mathrm{K}^{+}}$ around the vicinity of the rod ($R\sim5\,\mathrm{nm}$), with a subsequent reduction of the net negative charge at longer distances (thick lines in gray). 
The amount of condensation is larger for the rod with $\rho_{_{\mathrm{1D}}}=$ $-5\,e/\mathrm{nm}$ than for one with $\rho_{_{\mathrm{1D}}}=$ $-2\,e/\mathrm{nm}$.
In the case of the lowest charge density ($-1$ $e/\mathrm{nm}$ corresponding $\xi=0.7$ ), condensation is not observed as predicted by the CIC theory (Fig.~\ref{fig:rod_radial}c). 

The condition for condensation, $\xi>\nicefrac{1}{\lvert z\rvert}$, also implies that the charged rod must be surrounded to a greater degree by $\textrm{Mg}^{2+}$ than $\textrm{K}^{+}$.
This prediction was also borne out in the simulations.
In Fig.~\ref{fig:rod_radial} (d and e), \Mg{} ions condense in the immediate vicinity of the rods taking over the role of \K{} in the charge renormalization.
The replacement of \K{} by \Mg{} in the vicinity of the rod is expected based on the counterion release mechanism (leads to an entropy of the whole system increase), which is another well-known consequence of the CIC theory.
The condensation of cations, and the subsequent net reduction of negative charge on the rod, are more efficient in the presence of \Mg{} than when $z=1$ (compare panels (a) and (d) in Fig.~\ref{fig:rod_radial}). 
Since the condensation condition for \Mg{} is $\xi>0.5$, a distinct increase of $Q_{\mathrm{Mg}^{2+}}$ is visible even at the lowest density case, $\rho_{_{\mathrm{1D}}}=$ $-1$ $e/\mathrm{nm}$ ($\xi=0.7$ ) albeit the extent of condensation is not as great as found in the  two rod systems with higher density (Fig.~\ref{fig:rod_radial}f). 

\begin{figure}[hbt!]
\centering
\includegraphics[width=0.6\textwidth]{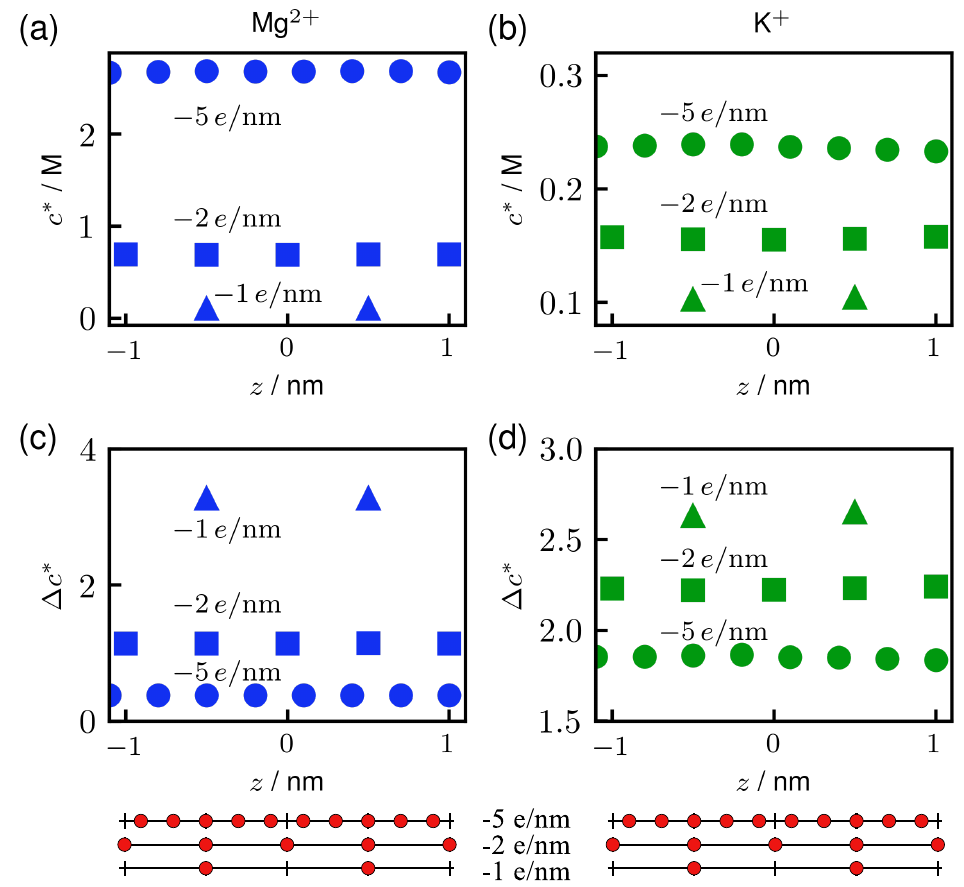}
\caption{
Condensation of \Mg{} (a, c) and \K{} (b, d) for the three rods with differing charge density, $\rho_{_{\mathrm{1D}}}$, (Fig.~\ref{fig:setup}a) in 5 mM \Mg{} and 50 mM \K{} buffer.
\textbf{(a, b)} Local ion concentrations of\textbf{ }(a) $\textrm{Mg}^{2+}$ and (b) $\textrm{K}^{+}$.
\textbf{(c, d)} Normalized fluctuations (Eq.~\ref{eq:fluctlocalconc}) of the local ion concentrations of (c) \Mg{} and (d) \K{} at the same condition as in (a) and (b).
The positions of negative charges on the three rods are indicated at the bottom. 
In order to avoid end effects, only the middle region of the rod ($-1.1\,\text{nm}\le z\le1.1\,\text{nm}$) was used in the analysis.
}
\label{fig:rod_z}
\end{figure}

The results quoted above using the control simulations are entirely in accord with the CIC theory predictions.
The extremely close correspondence between the CIC theory and simulations for charged rods motivated us to make several assessments of ion condensation phenomena in ribozyme folding, in which we use the same explicit-ion model.
In order to make quantitative comparisons between the rod and RNA folding simulations, we need to devise a quantity to precisely measure the ion condensation.
The radial distributions in cylindrical coordinates are not suitable for flexible RNA chains, for which the persistence length (1 $\sim$ 2 nm for a single strand) and the Debye screening length (1.3 nm in the simulation conditions) are similar.
With this in mind, we defined local ion concentrations ($c^{\ast}$) and the associated fluctuations (Eqs.~\ref{eq:localconc}-\ref{eq:fluctlocalconc}) in the spherical coordinate for each point charge. 
The results of the rod simulations in terms of $c^{\ast}$ are shown in Fig.~\ref{fig:rod_z} (a and b) for the three rod charge densities in the presence of 5 mM MgCl$_{2}$ and 50 mM KCl.
It is clear that the density of condensed ions around the charges in the rod is uniform; there is no position dependence along the rod.
From panels (c) and (d) in Fig.~\ref{fig:rod_z}, the fluctuation in $c^{\ast}$ is the lowest in the case of \Mg{} condensed for the highest charge density rod. 
Both lowering the linear charge density and reducing the valence of cations (\Mg{} $\rightarrow$ \K{}) make the condensed state less favorable, thus increasing the magnitude of charge fluctuations. 
These quantitative descriptions for the charged rods set the stage for analyzing the completely contrasting behavior in ion-ribozyme interactions.

\subsection*{[\Mg{}]-dependent folding of {\it Azoarcus} ribozyme}
\begin{figure}[hbt!]
\centering
\includegraphics[scale=1.0]{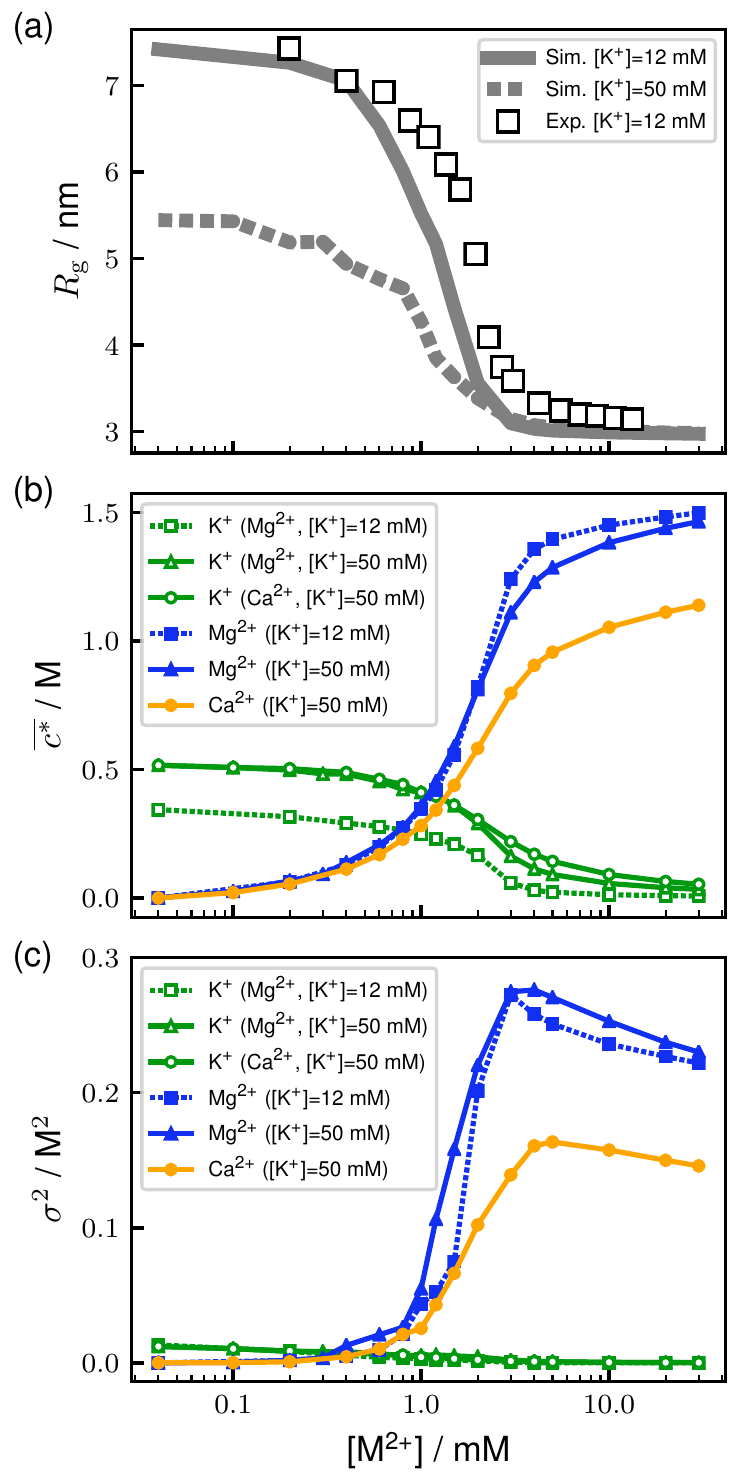}
\caption{
Ribozyme folding and ion condensation as a function of \Mg{} concentration. \textbf{(a)} Average radius of gyration ($R_{g}$) as a function of the concentration of \Mg{}, that indicates the global compaction around the midpoint of {[}\Mg{}{]} $\sim1.5$ mM.
Squares are experimentally measured $R_{\textrm{g}}$ in a Tris buffer containing 12 mM monovalent salt \cite{Behrouzi12Cell}.
\textbf{(b)} [\Mg{}]-dependence of local ion concentrations averaged over the nucleotide sequence, $\overline{c^{\ast}}$, for \K{}, \Mg{}, and \Ca{}.
\textbf{(c)} Variances in the concentrations along the RNA sequence, $\sigma^{2}$. 
}
\label{fig:salt_average}
\end{figure}

Before proceeding to examine the roles of ions in RNA folding, we summarize briefly the [\Mg{}]-dependent folding of the \textit{Azoarcus} ribozyme, which has been investigated in a previous study using the elaborate version of the TIS model \cite{Denesyuk15NatChem}.
The folding process of the ribozyme (Fig.~\ref{fig:setup}) strongly depends on the concentration of $\textrm{Mg}^{2+}$.
In Fig.~\ref{fig:salt_average}a, the dependence of the radius of gyration ($R_{\textrm{g}}$) on [\Mg{}] is shown in gray lines for two different conditions of monovalent salt concentration, [\K{}] = 12 and 50 mM.
The global compaction of the RNA appears as a single transition in $R_{\textrm{g}}$ occurring at  $C_{\textrm{m}} \approx1.5$ mM, where we designate $C_{\textrm{m}}$ as the midpoint concentration of [\Mg{}] separating the folded and unfolded states.
Comparison of 12 mM \K{} simulation data and SAXS experiments at the equivalent condition \cite{Behrouzi12Cell} (squares in Fig.~\ref{fig:salt_average}) shows excellent agreement, thus establishing that our model captures monovalent- and divalent-ion dependence of folding of {\it Azoarcus} ribozyme.
In the absence of \Mg{}, the key tertiary interactions (marked in red in Fig.~\ref{fig:setup}c) are disrupted and only some of the helical domains, P2, P4, P5, and P8 are stably formed. 
As [\Mg{}] increases, other helices and tertiary interactions form in multiple steps. 
At [\Mg{}]~$\sim 1$~mM, tertiary interactions, TH, G site, and SE form, followed by folding of other peripheral interactions, TL2-TR8 and TL9-TR5, at [\Mg{}]~$>\,2$~mM. The formations of these tertiary interactions lead to a compact intermediate state with $\left.\langle R_{\textrm{g}} \rangle\right.\sim 3.5$~nm.
Some interesting cooperativity and anti-cooperativity, such as  cooperative formation of P3 helix and SE, were found and discussed in the previous study \cite{Denesyuk15NatChem}.
The complete native structure, including the active site, is spontaneously formed at [\Mg{}]~$>\,4$~mM.
In subsequent parts of this paper, we report detailed analyses of ion distributions around the RNA. 
Comparisons with the interaction of divalent ions with highly charged rods illustrate the importance of ribozyme architecture on ion--RNA interactions.

\subsection*{Radial profiles of ion condensation}
\begin{figure}[hbt!]
\centering
\includegraphics[width=0.8\textwidth]{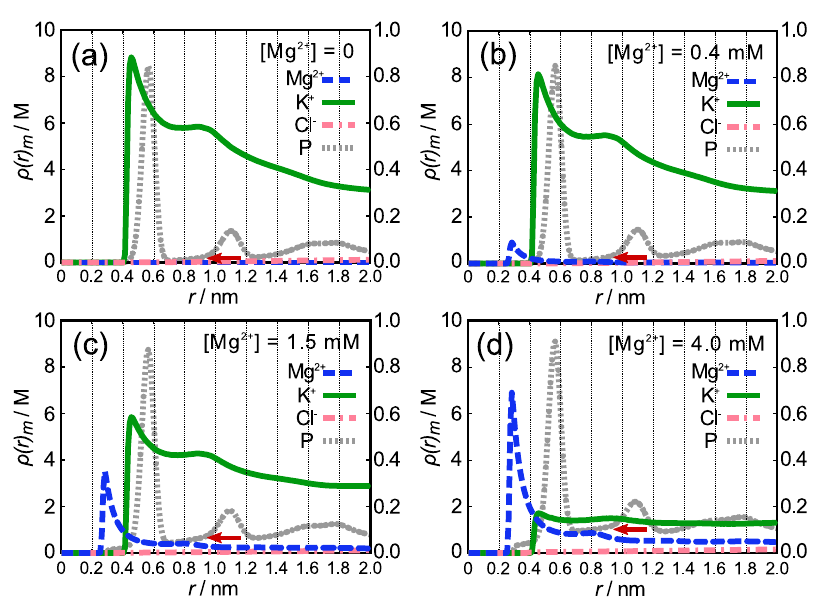} 
\caption{
Radial profiles of ion distributions around the ribozyme, $\rho(r)$, for $\textrm{Mg}^{2+}$, $\textrm{K}^{+}$, $\textrm{Cl}^{-}$, and phosphate (P) at various \Mg{} concentrations: (a) 0, (b) 0.4 mM, (c) 1.5 mM, and (d) 4.0 mM. 
The concentration of \K{} is 50 mM.
An appropriate number of Cl$^{-}$ were added to the solution for the electroneutrality.
Note that the scale for $\rho_{_{\text{Mg}^{2+}}}$ and $\rho_{_{\textrm{P}}}$ (left axis) is 10 times greater than $\rho_{_{\textrm{K}^{+}}}$ and $\rho_{_{\textrm{Cl}^{-}}}$ (right axis).
It should be pointed out that the density profile for \K{} decreases substantially as {[}\Mg{}{]} increases.
The red arrows indicate the amplitude of non-neighboring phosphates. 
The results in 12 mM KCl are shown in Fig.~S1 in the SI.
The density profiles for \Mg{} and P are roughly independent of the $\textrm{K}^{+}$ concentration.
}
\label{fig:Radial}
\end{figure}

We calculated the density
profiles of each ion species around phosphate, $\rho_{i}(r)$, where
$r$ is the distance from the phosphate of the $i$\textsuperscript{th} nucleotide. We also computed the profiles of phosphate--phosphate distance, which can be used to trace the compaction of the ribozyme.
The ensemble-averaged profiles, for given \Mg{} concentrations, are shown in Fig.~\ref{fig:Radial}.
From the profiles of the phosphate (gray dotted lines in Fig.~\ref{fig:Radial}), we can track the folding of the ribozyme as follows. 
The peak at $r\approx0.55$ nm is roughly at the distance between two consecutive phosphates, and the smaller second peak corresponds to the location of the second nearest-neighbor phosphates.
The amplitudes of these two peaks do not show significant change as {[}\Mg{}{]} increases.
However, the folding process can be discerned by the increase in $\rho_{_{\mathrm{P}}}(r)$ in the range of 0.7 $<r<$ 0.9 nm (red arrows in Fig.~\ref{fig:Radial}).
At low {[}\Mg{}{]}, there is practically no density in $\rho$ in this range ($\rho(r)\approx0$).
As {[}\Mg{}{]} increases, we find that the amplitude increases, which shows that \Mg{} ions facilitate the approach of non-neighboring phosphates.
The increase in $\rho(r)$ in this range coincides with the compaction of the ribozyme and subsequent folding as shown in Fig.~\ref{fig:salt_average}(a).

In the absence of \Mg{}, \K{} cations are condensed around the ribozyme, regardless of the monovalent salt concentration, {[}KCl{]} = 12 mM or 50 mM (Figs.~\ref{fig:Radial}a and S1a). 
Under these conditions with [\Mg{}] $=0$, all tertiary interactions are disrupted but helices, except P3 and P7, form secondary structures (Fig.~\ref{fig:setup}c).
Thus, \K{} cations must be condensed around the vicinity of phosphates to balance the electrostatic repulsion between negative charges on the phosphates. The distance to condensed \K{}, $r\approx0.4$ nm, is closer than the neighboring phosphate.
In the presence of \Mg{}, $\rho(r)$ for \K{} decreases dramatically as the \Mg{} concentration increases (Fig.~\ref{fig:Radial} b-d).
This shows, rather vividly, that \K{} ions condensed around RNA are replaced with $\textrm{Mg}^{2+}$ ions because divalent cations screen negative charges on the RNA more efficiently than monovalent cations.
By replacing two \K{} ions by one $\mbox{Mg}^{2+}$, there is a net gain in translational entropy of the ions. This is the counterion release mechanism predicted by the CIC theory.

\subsection*{Fingerprint of local ion concentration $c^{\ast}_{i}$}
In order to examine the relationship between ion condensation and \Mg{}-driven RNA folding, we calculated the local ion concentration of the $i$\textsuperscript{th} nucleotide $c_{i}^{\ast}$ (Eq.~\ref{eq:localconc}).
The cutoff distance to define the local volume is chosen to be the Bjerrum length, $r_{c}=0.73$ nm.
The range of this distance covers the peaks in the radial distributions of both \K{} and \Mg{} (Fig.~\ref{fig:Radial}).
Accordingly, $c_{i}^{\ast}$ is expected to capture the amount of ion atmosphere around each nucleotide, and such a fingerprint is dependent on the architecture of the folded state of the RNA.
The overall ion condensation of the entire molecule, $\overline{c^{\ast}}$, is calculated as the average of the local ion concentration along the RNA sequence (Eq.~\ref{eq:averagelocalconc}, see Models and Methods).

The changes in $\overline{c^{\ast}}$ as a function of [\Mg{}] is shown in Fig.~\ref{fig:salt_average}(b).
As noted previously, more $\textrm{K}^{+}$ cations condense at lower \Mg{} than at higher \Mg{} concentrations.
Indeed, $\overline{c_{\textrm{K}^{+}}^{\ast}}$ monotonically decreases as [\Mg{}] increases (green lines).
The folding midpoint of \Mg{} concentration is around 1.5 mM. Above the midpoint, \Mg{} ions are more dominantly found in the vicinity of the RNA than \K{}.
$\overline{c_{\textrm{Mg}^{2+}}^{\ast}}$ does not vary significantly as a function of \K{} concentrations for the entire range of [\Mg{}] (Fig.~\ref{fig:salt_average}b, solid line for [\K{}] = 50 mM and dotted line for 12 mM).
The sharp increase in $\overline{c_{\textrm{Mg}^{2+}}^{\ast}}$ is well correlated with the global compaction of the ribozyme, that is evident as the decrease in $R_{\mathrm{g}}$ in Fig.~\ref{fig:salt_average}(a).

\subsection*{Ion condensation is site-specific}
In Fig.~\ref{fig:salt_average}c, we show variances, $\sigma^{2}$, of the local ion concentration over the RNA sequence (Eq.~\ref{eq:deviationlocalconc}).
This quantity indicates how much of each ion species are localized at particular positions along the sequence at each solution condition.
As \Mg{} concentration increases, $\sigma^{2}$ for \K{}monotonically decreases, although it is much smaller compared to divalent cations for the entire range of {[}\Mg{}{]}. 
This means that \Mg{} ions tend to be localized at specific positions more than \K{} ions.
In particular, $\sigma^{2}$ for \Mg{} sharply increases reaching a maximum around {[}\Mg{}{]} $\approx 3$ mM. 
This is because \Mg{} ions are preferentially localized to specific places, which are, surprisingly,  determined by the tertiary structure of the ribozyme. 
Beyond the midpoint $C_{\mathrm{m}}$, the value of $\sigma^{2}$ decreases modestly because the specific sites of higher phosphate densities are occupied by existing localized \Mg{}. The newly added \Mg{} ions start contributing to lessening the gap between specific and other non-specific sites.

\begin{figure}[hbt!]
\centering
\includegraphics[width=0.95\textwidth]{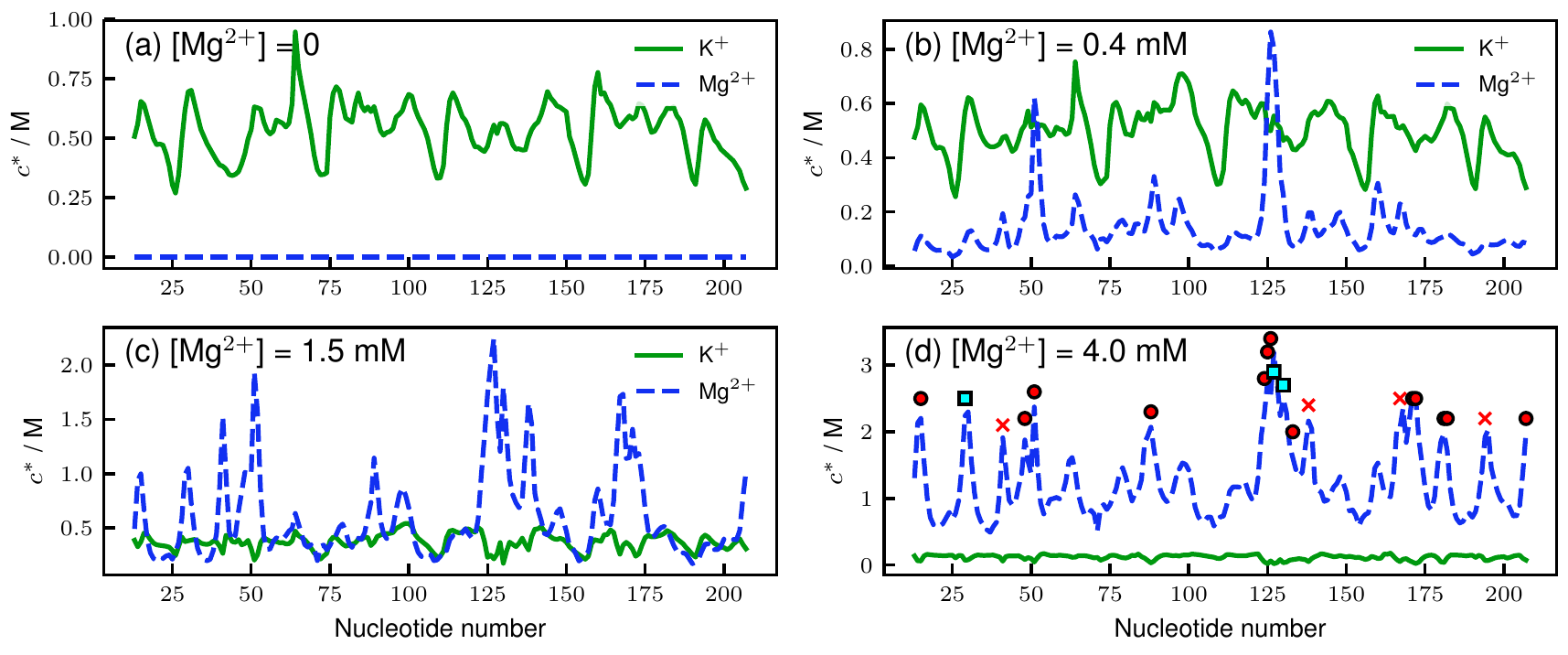}
\caption{
Fingerprint of ion coordination expressed as local ion concentrations, $c_{i}^{\ast}$ ($i$ is the nucleotide position), for $\textrm{Mg}^{2+}$ (blue) and $\textrm{K}^{+}$ (green).
The local ion concentrations are calculated as the molarity of ions in the vicinity of each nucleotide ($r\leq\ell_{\mathrm{B}}=0.73\,\textrm{nm}$).
The peaks with varying heights at different nucleotides positions as ribozyme folds reflect the architecture of the folded state. 
Most importantly, $c^{\ast}$ is not uniform but varies dramatically along the nucleotides, which is very different from what is found for charged rods (see Fig.~\ref{fig:rod_z}a).
In the panel (d), positions of the \Mg{}-binding sites found in the crystal structure \cite{AzoXray04N} are indicated by red circles (direct contact; nucleotides 15, 48, 51, 88, 124-126, 133, 171-172, 181-182, 207) and cyan squares (contact mediated by water molecules; nucleotides 29, 127, 130). The positions with high $c^{\ast}$ in the simulations but no \Mg{} in the crystal structure are marked by red crosses (nucleotides 41, 138, 167, 194).
The results at 50 mM \K{} are shown in Fig.~S2 in the SI. 
}
\label{fig:nt-dependence}
\end{figure}

Fig.~\ref{fig:nt-dependence} shows the distributions of the local ion concentration along the RNA sequence.
This figure shows that all non-zero values of \Mg{} condensation is highly specific. 
The extent of condensation depends on the nucleotide. 
Even at the lowest [\Mg{}] $=0.4$ mM (much lower than $C_{\mathrm{m}}$) there is a significant density of \Mg{} around $i\approx125$, which covers the triple helix (TH) of {\it Azoarcus} ribozyme. 
Thus, we conclude that \Mg{} binding is highly specific. 
More importantly, charge neutralization resulting in $\beta<1$, occurs at the lowest [\Mg{}] not uniformly, as predicted by the CIC theory, but does so in a manner that initiates folding. 
This would mean that different regions of this ribozyme order at somewhat distinct \Mg{} concentrations, as illustrated previously (see Fig.~2b in \cite{Denesyuk15NatChem}).
In other words, the midpoint concentration, $C_{\mathrm{m}}$, of the transition measured by $R_{\textrm{g}}$ or other parameters is the (appropriately-weighted) mean of such individual [\Mg{}] at which distinct structural elements achieve the native conformation.

\begin{figure}[hbt!]
\centering
\includegraphics[width=0.6\textwidth]{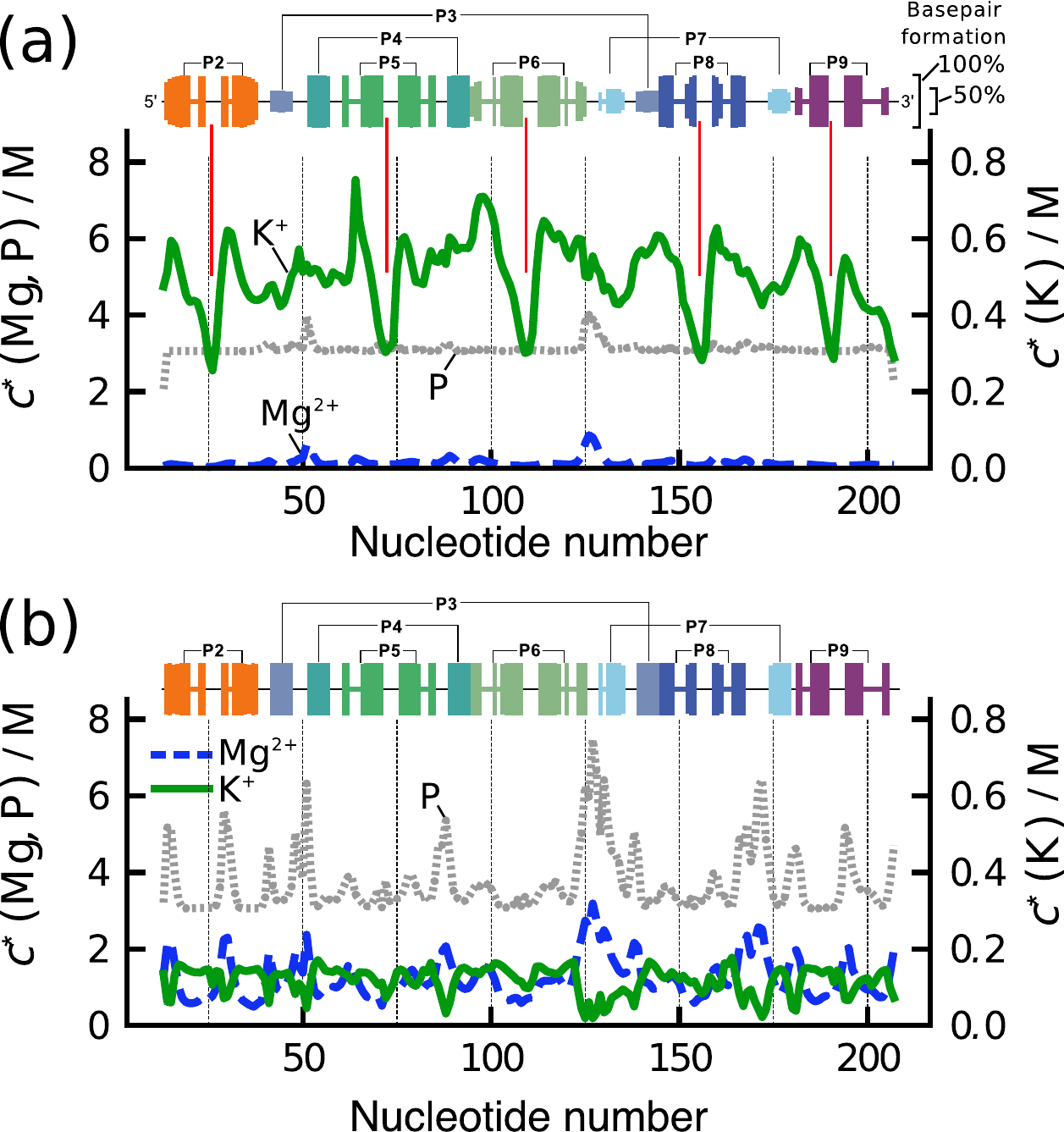}
\caption{
Relationship between \K{} condensation and residual structure.
The fraction of secondary structure formation is shown on top as bars for each nucleotide. \textbf{(a)} At 0.4 mM \Mg{}, the profile of condensed \K{} (green) has several characteristic minima that correspond to loops in the stem-loop structures of P2, P5, P6, P8 and P9 (indicated by vertical red lines). 
See Fig.~\ref{fig:setup} for the structures.
\textbf{(b)} At 4.0 mM \Mg{}, the profile of condensed \Mg{} correlates with the profile of phosphate (gray dashed). Such a correlation does not exist at low \Mg{} concentration. The secondary structures (various colored bars on top) are nearly fully formed.
}
\label{fig:nt-dependence-detail}
\end{figure}

In the absence of or at low \Mg{} concentrations, the amount of condensed \K{} depends significantly on positions along the RNA sequence. At low \Mg{} concentrations, the ribozyme is unfolded with some helices intact.
To see if there is any relationship between \K{} condensation and such residual structure, we compare the distribution of $c^{\ast}$ with the fraction of secondary structure formation in Fig.~\ref{fig:nt-dependence-detail}(a).
The profile of condensed \K{} (green in Fig.~\ref{fig:nt-dependence-detail}a) has several characteristic minima that correspond to loops in the stem-loop structures of P2, P5, P6, P8, and P9.
This indicates that the distribution of \K{} reflects the shape of RNA that is mostly due to residual secondary structures at low {[}\Mg{}{]}.

At high \Mg{} concentration ({[}\Mg{}{]} $>1.5$ mM), the localization of condensed \Mg{} ions depends on the tertiary structures of the RNA (Fig.~\ref{fig:nt-dependence}c and d).
The local concentration of \Mg{} correlates with the local density of phosphates (compare blue and gray lines in Fig.~\ref{fig:nt-dependence-detail}b).
Under such conditions, all the native secondary structures are formed with almost unit probability, but we do not find any secondary-structure dependence of condensed \Mg{}. 

\begin{figure}[hbt!]
\centering
\includegraphics[width=0.6\textwidth]{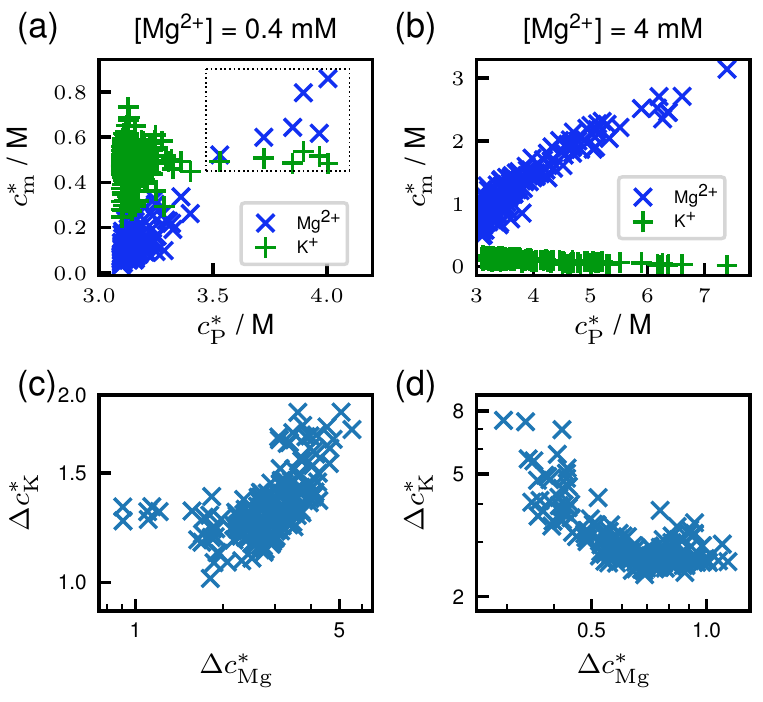}
\caption{
\textbf{(a, b) }Correlation between local ion concentrations of phosphate ($c_{\mathrm{P}}^{\ast}$) and the cations ($c_{m}^{\ast}$) where $m$ is either \K{} or \Mg{}.
Each data point represents one nucleotide. There are 195 such data points in each panel. The solution conditions are (a) {[}\Mg{}{]} = 0.4 mM, and (b) 4 mM, both in 50 mM KCl. 
In (a), the data points at $c_{\mathrm{P}}^{\ast}>3.5\,\mathrm{M}$ correspond nucleotides 50,
51, and 124-127 (inside the dotted rectangle), that are involving the tertiary helix. \textbf{(c, d) }Correlations of fluctuations,
$\Delta c^{\ast}$, between \Mg{} and \K{} at (c) [\Mg{}] = 0.4 mM and (d) 4.0 mM in 50 mM KCl.
}
\label{fig:cor_local}
\end{figure}

The relationship between ion condensation and local density of phosphates is vividly illustrated in Fig.~\ref{fig:cor_local}(a, b).
At low {[}\Mg{}{]}, the ribozyme is mostly unfolded, predominantly containing only secondary structures.
Because there are no stable tertiary interactions, $c_{\mathrm{P}}^{\ast}$ values are small (Fig.~\ref{fig:cor_local}a, horizontal axis. $c_{\mathrm{P}}^{\ast}\approx3\,\textrm{M}$). 
Accordingly, $c^{\ast}$ for \Mg{} are relatively low and \K{} are predominantly condensed.
However, there are six nucleotides that have higher $c^{\ast}$ for both P and \Mg{} (see the rectangle of $c_{\mathrm{P}}^{\ast}>3.5\,\mathrm{M}$ and $c_{\mathrm{Mg^{2+}}}^{\ast}>0.5\,\mathrm{M}$ in Fig.~\ref{fig:cor_local}a).
We identified all those nucleotides to be ones involving the central triple helix (TH) formation; nucleotides 50, 51 and from 124 to 127.
In accordance with the previous study \cite{Denesyuk15NatChem}, TH starts to form around {[}\Mg{}{]} = 0.4 mM ($< C_m$). The data in Fig.~\ref{fig:cor_local} indicates that there are \Mg{} ions specifically bound to the sites around TH.
In contrast, at high {[}\Mg{}{]}, the entire ribozyme is folded and $c_{\mathrm{P}}^{\ast}$ can be as high as $\sim$$7$ M depending on the local architecture (Fig.~\ref{fig:cor_local}b).
The amount of condensed \Mg{} is well correlated with the local concentration of phosphates.
There is a weak anti-correlation between \K{} condensation and local phosphate  concentration.

\subsection*{Fluctuations of condensed ions} 
RNA folding is a stochastic process driven by thermal fluctuations and the solution conditions.
Consequently, it would be interesting to see how these condensed ions fluctuate at different stages of folding.
In Fig.~S3, normalized fluctuations of the local ion concentrations, $\Delta c^{\ast}$ (Eq.~\ref{eq:fluctlocalconc}), for each nucleotide are shown for the several solution conditions.
Below or around the midpoint \Mg{} concentration, where the ribozyme folding is not completed, fluctuations of condensed \K{} ions and \Mg{} have similar position dependence. 
This correlation is more clearly seen in the logarithm plot of $\Delta c_{\mathrm{Mg}}^{\ast}$ vs $\Delta c_{\mathrm{K}}^{\ast}$ in Fig.~\ref{fig:cor_local}(c).
Comparing Figs.~\ref{fig:nt-dependence} and S3, we also find that nucleotide positions that have high $c^{\ast}$ values also have small $\Delta c^{\ast}$.
This indicates that the specific positions, where more ions are condensed, bind those ions also more tightly. Once the ions reach the ``equilibrium'' positions, they tend to be localized.

The correlation between the fluctuations of $\Delta c_{\mathrm{Mg}}^{\ast}$ and $\Delta c_{\mathrm{K}}^{\ast}$ changes dramatically after the midpoint, {[}\Mg{}{]} $>1.5$ mM (Fig.~S3d and e).
There is anti-correlation between $\log\Delta c_{\mathrm{Mg}}^{\ast}$ and $\log\Delta c_{\mathrm{K}}^{\ast}$ (Fig.~\ref{fig:cor_local}d). 
As discussed in previous sections, there are specific positions where more \Mg{} ions are condensed (Fig.~\ref{fig:nt-dependence}) beyond the midpoint of {[}\Mg{}{]}.
At these positions, there are smaller fluctuations in $c_{\mathrm{Mg}}^{\ast}$ (smaller $\Delta c_{\mathrm{Mg}}^{\ast}$) but higher fluctuations in $c_{\mathrm{K}}^{\ast}$ (larger $\Delta c_{\mathrm{K}}^{\ast}$), reflecting strong condensation of \Mg{}.

\subsection*{Ion condensation in Calcium-driven folding}
We have hitherto focused on the condensation of \K{} in the absence and presence of \Mg{}, and the condensation of \Mg{} itself.
It is known that \Ca{} could also facilitate $Azoarcus$ ribozyme folding although the folded state is not catalytically active and is modestly less compact than it is in the presence of \Mg{} \cite{Behrouzi12Cell,Denesyuk15NatChem}.
We repeated the same analyses in the case of \Ca{}-dependent folding instead of \Mg{}.
Both average local concentration ($\overline{c^{\ast}}$) and its variance ($\sigma^{2}$) show similar trends as the case of \Mg{} (Fig.~\ref{fig:salt_average}b and c).
At higher solution concentration ({[}Mg/\Ca{}{]} $\gtrsim1$ mM), the amount of condensation given by $\overline{c^{\ast}}$ is slightly less in the case of \Ca{}. 

Fig.~S4 shows the sequence dependence of the local concentration
($c^{\ast}$) and fluctuation ($\Delta c^{\ast}$) in the case of
\Ca{}. Comparing them with Fig.~\ref{fig:nt-dependence} and
S2, the condensation profiles of \Ca{} and \K{} bear remarkable
resemblances to the \Mg{} case. At the highest concentration
([Mg/\Ca{}] $=30$ mM), the local concentration of \Ca{}
has the same peak positions as \Mg{}, but has $\sim$20 \% lower
values. Comparison of the fingerprint of ion concentration profiles between \Mg{}
and \Ca{} clearly illustrates the efficacy of \Mg{} of creating a compact and functionally competent ribozyme.

\section*{Discussion}

The central result of this study is that the nature of ion condensation onto
{\it Azoarcus} ribozyme differs drastically from theoretical predictions
for regularly shaped polyanions. For example, the CIC theory shows that ions condense uniformly
onto highly charged rods. The positively charged ions are localized at the positions of the charges on the rod. In sharp contrast, in {\it Azoarcus} ribozyme, and presumably in other RNA molecules as well, the condensation is highly non-uniform and nucleotide specific, and critically depends on the structure of the RNA. We note in passing that the prevailing view in much of the RNA literature is that most of the ions are uniformly condensed onto RNA although crystal structures have shown that there are site-specific localizations of \Mg{}. A natural explanation of our finding is that the electrostatic potential around the phosphate groups is non-uniform, with the folded structure with fully formed tertiary interactions containing pockets of highly negative potentials. 
Consequently, ions are specifically drawn, with higher probability, to regions of large negative electrostatic potential with smaller ions approaching these regions easily. Indeed, this was first shown in our previous study using coarse-grained and atomic-detailed simulations (see Fig.~S11 in \cite{Denesyuk15NatChem}).
However, what is surprising in this study is that the ions sense these regions even at extremely low divalent ion concentrations when there is no tertiary interaction present. 

\begin{figure}[hbt!]
\centering
\includegraphics[width=0.6\textwidth]{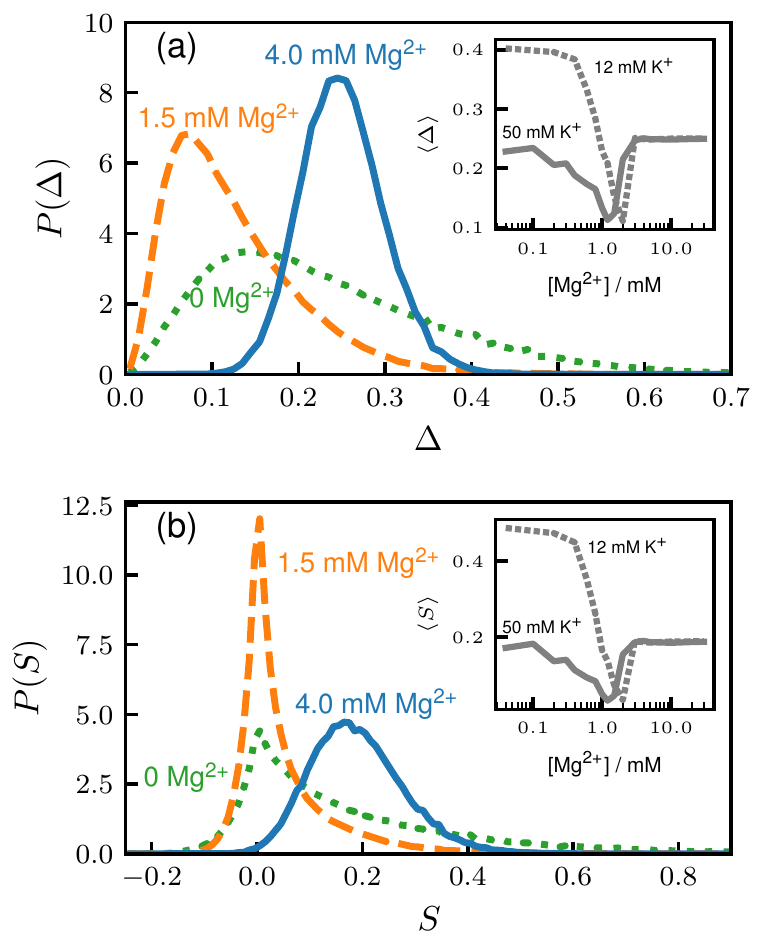}
\caption{
Probability distributions of shape parameters, (a) asphericity $\Delta$ and (b) prolateness $S$, for the {\it Azoarcus} ribozyme at three \Mg{} concentrations, 0 mM (dotted green), 1.5 mM (dashed orange), and 4 mM (solid blue). If the shape is a perfect sphere then $\Delta=S=0$, whereas a perfect rod corresponds to $\Delta=S=1$.
The insets show the average values of $\left<\Delta\right>$ and $\left<S\right>$ as a function of [\Mg{}] at [\K{}] = 12 mM (dotted line) and 50 mM (solid line). 
The values of $\left<\Delta\right>$ and $\left<S\right>$ in the folded states are 0.24 and 0.17, respectively.
Globally the shape of this ribozyme corresponds to a prolate ellipsoid.
}
\label{fig:shape}
\end{figure}

\subsection*{Link to the shape fluctuations in the ribozyme}
The non-uniform and site-specific ion-RNA interactions can be succinctly illustrated by considering the ion concentration-dependent changes in the shape parameters, asphericity $\Delta$ ($0\le\Delta\le1$) and prolateness $S$ ($-0.25\le S\le2$) \cite{Aronovitz86JP,Dima04JPCB}, which are readily computed from the inertia tensor that is related to the square of the radius of gyration.
For reference, it is worth pointing out both $\Delta$ and $S$ are unity for rods, whereas $\Delta=S=0$ for spheres.
For long self-avoiding polymers, numerical results show that $\Delta=0.55$ and $S=0.92$ (a prolate ellipsoid) \cite{Honeycutt89JCP}.
In contrast to the values of the shape parameters for rods, the changes in $\left<\Delta\right>$ and $\left<S\right>$ as a function of [\Mg{}] for the ribozyme, at the two values of \K{} concentration, exhibit unusual non-monotonic dependence as [\Mg{}] is varied (inset in Fig.~\ref{fig:shape}), which is not reflected in $R_{\mathrm{g}}$.
Fig.~\ref{fig:shape} also shows the distributions $P(\Delta)$ and $P(S)$ at three \Mg{} concentrations. 
Surprisingly, the widths of $P(\Delta)$ and $P(S)$, which depend non-monotonically on [\Mg{}], are relatively large.
Therefore, shape fluctuations in the ribozyme are substantial.
Two comments are worth making: 
(1) The values of $\Delta$ and $S$ depend greatly on the \K{} concentration, which is also reflected in the $R_{\mathrm{g}}$ as a function of \Mg{} concentration.
(2) It is is interesting that the minima in $\left<\Delta\right>$ and $\left<S\right>$ are at [\Mg{}] $\approx1.5$ mM, the midpoint of the folding transition, where the ribozyme almost resembles a sphere. At high and low \Mg{} concentrations, the values of $\left<\Delta\right>$ show that globally the ribozyme may be pictured as a prolate ellipsoid. It also implies that this ribozyme and others are not as densely packed as proteins \cite{Hyeon06JCP}.
The irregular shape leads to the high specificity of ion--RNA interactions. It is likely that this finding would apply to any RNA molecules whose shape has grooves and whose surface are irregular.
It is most interesting that there is a calculable fingerprint for ion-RNA interactions that is a reflection of the folded structure.
A clear implication is that, even at very low ion concentration, binding is highly specific. 
This interpretation differs from the usual assertion that the stability of RNA is determined largely by diffuse divalent cations \cite{Soto07Biochem}. 

\subsection*{Correlation between ions at different RNA sites}
We computed the correlations of \Mg{}/\K{} binding to two different nucleotides (Fig.~S5).
The correlations for a pair of nucleotides ($i$, $j$) was calculated as the Pearson's coefficient of the number of \Mg{}/\K{} ions bound to site $i$ and site $j$ for the ensemble at a given \Mg{} and \K{} concentrations.
Overall, we did not find strong correlations involving \Mg{} or \K{}, except that some signatures appear as consequences of formations of secondary and tertiary structures.
For example, there are strong correlations near the diagonal region because of helix domains P5, P6, and P8, as indicated by the red circles (Fig.~S5a).
In addition, there are high correlations at the interfaces of P3/P7 (see Figure~4 of Ref.~\cite{Denesyuk15NatChem}), and P2/P8.
The observed high correlations arise because these nucleotides are close in space and share the same \Mg{} ion when the secondary/tertiary interactions form.
Interestingly, at the midpoint concentration [\Mg{}] $\sim1.5$~mM, there are weak anti-correlations between P2 and P6a, and P2 and J8/7 (blue circles).
Helix P2 is located at a peripheral part of the ribozyme, whereas P6a and J7/8 are close to the core region.
These anti-correlations are a consequence of anti-cooperative folding of tertiary interactions, arising from topological frustration \cite{Denesyuk15NatChem}.
Around the midpoint concentration of [\Mg{}], there are still not enough amount of \Mg{} ions to simultaneously satisfy all the folding elements.
Thus, \Mg{} binding to P2 reduces the probability of \Mg{} binding to the core region.
The correlation maps between \K{}-\K{} ions are less instructive.

\subsection*{Flexible polyelectrolytes}
Just as is the case for RNA, which does adopt a well-defined folded structure at high ion concentrations, the shapes of flexible PEs are also highly variable.
Therefore, it follows that the CIC theory should not be quantitative in predicting the nature of ion condensation.
Indeed, early simulations showed \cite{Lee01Macromolecules} that ion condensation is non-uniform in the presence of divalent and trivalent cations for PEs in poor solvents.
Indeed, in certain cases, the divalent cations and the monomers are packed as in a crystalline arrangement, which would be inconsistent with the CIC theory.
Using theory, Muthukumar \cite{Muthukumar04JCP} has shown that, even in monovalent salts ($z$=1), the condensation process in flexible PEs is non-uniform and could be understood by the variations in the local dielectric mismatches.
A more recent simulation study \cite{Chremos16SoftMatter} has also arrived at the conclusion that the condensation of monovalent ions depends crucially on the shape of the PE.
In particular, they showed that non-uniform condensation is determined by the molecular topology. Thus, the conclusion that shapes determine ion condensation holds for both synthetic PEs and RNA.
The latter has additional features (for example stacking interactions and hydrogen bonding), which make RNA different from a synthetic.
In particular, the ion-induced shape changes in PEs are not as spectacular as they are in RNA.  

\section*{Conclusion}

The Oosawa-Manning theory of ion condensation, which assumes that there is an equilibrium between bound and free ions, is the basis of our understanding of many thermodynamics properties of highly charged polyelectrolytes. 
As shown multiple times previously and in the simulations here, the predictions of the CIC theory are quantitatively correct if the charged polyelectrolytes adopt regular shapes, such as rods and spheres.
This is not the case in RNA, which although is a polyanion, does not adopt a simple shape but undergoes large shape transitions as it folds.
Our simulations of {\it Azoarcus} ribozyme in the presence of \Mg{} or \Ca{} in a buffer containing \K{} ions show that ion binding is highly specific.
Surprisingly, the specificity of binding depends on the folded structure of the RNA. 
In other words, in this ribozyme with a complex fold, the charge neutralization of phosphates does not occur uniformly as predicted by the CIC theory for rod-like macroions.
There are specific regions where condensation occurs even at very low concentrations of the divalent cations.
Although we have arrived at this unexpected finding using simulations of one ribozyme, we expect that the main conclusion should be applicable for other RNA molecules as well. 

We conclude with the following additional comments.
(1) We find, perhaps not surprisingly, that monovalent ions unbind (termed as counterion release process) when RNA is titrated with divalent ions (Fig. 6).
This validates one of the predictions of the CIC theory. The release of \K{}, however, occurs from specific nucleotide locations on the RNA, which is not anticipated in the CIC theory. 
(2) It is unclear to us if our findings, especially the exquisite specificity of the predicted nucleotide-dependent binding of ions to RNA, could be tested in experiments because of the highly correlated and many-body nature of ion condensation.
Multicolor FRET experiments may be used to label regions of high affinity for divalent cations.
The conformational changes as a consequence of ion binding and unbinding at these locations could be used, most likely in conjunction with simulations, to determine the architecture-dependent interaction of ions with RNA.
(3) Our results show that charge renormalization of phosphates occurs in a highly heterogeneous manner and not by the diffuse binding picture of ion-RNA interactions.
The fingerprints of ion localizations show that the extent of charge neutralization depends on the nucleotide position -- finding that has to be incorporated in calculating free energy changes of RNA as it folds.
(4) The clear implication of our results is that, even when using coarse-grained models, ions (especially those with valence greater than unity) should be modeled explicitly to reflect their sizes and charge densities.
It is possible that electrostatic interactions due to monovalent ions can be accounted for by using effective interactions \cite{Denesyuk18JPCB}.
However, in order to obtain accurate thermodynamics of RNA folding, divalent ions have to be explicitly modeled \cite{Denesyuk15NatChem,Hayes15PRL}.

\section*{Author Contributions}
NH, NAD, and DT designed the research; NH and NAD performed all the simulations; NH, NAD, and DT analyzed the data; NH and DT wrote the article.

\section*{Acknowledgments}
We thank Dr. Hung T. Nguyen for useful discussions. This work was supported by a grant from the
National Science Foundation (CHE 16-36424) and the Welch Foundation (F-0019) through the Collie--Welch Chair.

\bibliography{AzoCondensation}

\begin{thebibliography}{31}
\providecommand{\url}[1]{\texttt{#1}}
\providecommand{\urlprefix}{ }

\bibitem[Sun et~al.({2017})Sun, Zhang, and Chen]{Sun17ARB}
Sun, L.-Z., D.~Zhang, and S.-J. Chen, {2017}.
\newblock Theory and modeling of {RNA} structure and interactions with metal
  ions and small molecules.
\newblock \emph{Annu. Rev. Biophys.} {46}:{227--246}.

\bibitem[Yu and Chen({2018})]{Tao18BJ}
Yu, T., and S.-J. Chen, {2018}.
\newblock Hexahydrated {Mg}$^{2+}$ binding and outer-shell dehydration on {RNA}
  surface.
\newblock \emph{Biophys. J.} {114}:{1274--1284}.

\bibitem[Pollack({2011})]{Pollack11Biopolymers}
Pollack, L., {2011}.
\newblock Time resolved {SAXS} and {RNA} folding.
\newblock \emph{{Biopolymers}} {95}:{543--549}.

\bibitem[Kirmizialtin et~al.({2012})Kirmizialtin, Pabit, Meisburger, Pollack,
  and Elber]{Kirmizialtin12BJ}
Kirmizialtin, S., S.~A. Pabit, S.~P. Meisburger, L.~Pollack, and R.~Elber,
  {2012}.
\newblock {RNA} and its ionic cloud: Solution scattering experiments and
  atomically detailed simulations.
\newblock \emph{Biophys. J.} {102}:{819--828}.

\bibitem[Denesyuk and Thirumalai(2013)]{Denesyuk13JPCB}
Denesyuk, N.~A., and D.~Thirumalai, 2013.
\newblock Coarse-grained model for predicting {RNA} folding thermodynamics.
\newblock \emph{J. Phys. Chem. B} 117:4901--4911.

\bibitem[Oosawa(1957)]{Oosawa57JPolymerSci}
Oosawa, F., 1957.
\newblock {A simple theory of thermodynamic properties of polyelectrolyte
  solutions}.
\newblock \emph{J. Polym. Sci.} 23:421--430.

\bibitem[Manning(1969)]{Manning69JCP1}
Manning, G.~S., 1969.
\newblock Limiting laws and counterion condensation in polyelectrolyte
  solutions. {I}. Colligative properties.
\newblock \emph{J. Chem. Phys.} 51:924--933.

\bibitem[Oosawa(1971)]{Oosawa1971Book}
Oosawa, F., 1971.
\newblock Polyelectrolytes.
\newblock Marcel Dekker.

\bibitem[Alexander et~al.(1984)Alexander, Chaikin, Grant, Morales, Pincus, and
  Hone]{Alexander84JCP}
Alexander, S., P.~Chaikin, P.~Grant, G.~Morales, P.~Pincus, and D.~Hone, 1984.
\newblock Charge renormalization, osmotic pressure, and bulk modulus of
  colloidal crystals: Theory.
\newblock \emph{J. Chem. Phys.} 80:5776--5781.

\bibitem[Heilman-Miller et~al.(2001{\natexlab{a}})Heilman-Miller, Thirumalai,
  and Woodson]{HeilmanMiller01JMBeq}
Heilman-Miller, S.~L., D.~Thirumalai, and S.~A. Woodson, 2001.
\newblock Role of counterion condensation in folding of the {T}etrahymena
  ribozyme {I}. Equilibrium stabilization by cations.
\newblock \emph{J. Mol. Biol.} 306:1157--1166.

\bibitem[Hyeon et~al.(2006)Hyeon, Dima, and Thirumalai]{Hyeon06JCP}
Hyeon, C., R.~I. Dima, and D.~Thirumalai, 2006.
\newblock Size, shape, and flexibility of {RNA} structures.
\newblock \emph{J. Chem. Phys.} 125:194905--10.

\bibitem[Denesyuk and Thirumalai(2015)]{Denesyuk15NatChem}
Denesyuk, N.~A., and D.~Thirumalai, 2015.
\newblock How do metal ions direct ribozyme folding?
\newblock \emph{Nature Chem.} 7:793--801.

\bibitem[Heilman-Miller et~al.(2001{\natexlab{b}})Heilman-Miller, Pan,
  Thirumalai, and Woodson]{HeilmanMiller01JMBkin}
Heilman-Miller, S.~L., J.~Pan, D.~Thirumalai, and S.~A. Woodson, 2001.
\newblock Role of counterion condensation in folding of the {T}etrahymena
  ribozyme {II}. Counterion-dependence of folding kinetics.
\newblock \emph{J. Mol. Biol.} 309:57--68.

\bibitem[Thirumalai et~al.(2001)Thirumalai, Lee, Woodson, and
  Klimov]{Thirumalai01AnnRevPhysChem}
Thirumalai, D., N.~Lee, S.~A. Woodson, and D.~Klimov, 2001.
\newblock Early events in {RNA} folding.
\newblock \emph{Annu. Rev. Phys. Chem} 52:751--762.

\bibitem[Adams et~al.(2004)Adams, Stahley, Kosek, Wang, and
  Strobel]{AzoXray04N}
Adams, P.~L., M.~R. Stahley, A.~B. Kosek, J.~Wang, and S.~A. Strobel, 2004.
\newblock Crystal structure of a self-splicing group {I} intron with both
  exons.
\newblock \emph{Nature} 430:45--50.

\bibitem[Gr{\o}nbech-Jensen et~al.(1997)Gr{\o}nbech-Jensen, Mashl, Bruinsma,
  and Gelbart]{GronbechJensen97PRL}
Gr{\o}nbech-Jensen, N., R.~J. Mashl, R.~F. Bruinsma, and W.~M. Gelbart, 1997.
\newblock Counterion-induced attraction between rigid polyelectrolytes.
\newblock \emph{Phys. Rev. Lett.} 78:2477--2480.

\bibitem[Borukhov et~al.(1997)Borukhov, Andelman, and Orland]{Borukhov97PRL}
Borukhov, I., D.~Andelman, and H.~Orland, 1997.
\newblock Steric effects in electrolytes: A modified {P}oisson--{B}oltzmann
  equation.
\newblock \emph{Phys. Rev. Lett.} 79:435.

\bibitem[Deserno et~al.(2000)Deserno, Holm, and May]{Deserno00Macromolecules}
Deserno, M., C.~Holm, and S.~May, 2000.
\newblock Fraction of condensed counterions around a charged rod: Comparison of
  {P}oisson--{B}oltzmann theory and computer simulations.
\newblock \emph{Macromolecules} 33:199--206.

\bibitem[Cha et~al.(2018)Cha, Ro, and Kim]{Cha18PRL}
Cha, M., S.~Ro, and Y.~W. Kim, 2018.
\newblock Rodlike counterions near charged cylinders: {C}ounterion condensation
  and intercylinder interaction.
\newblock \emph{Phys. Rev. Lett.} 121:058001.

\bibitem[Hyeon and Thirumalai(2005)]{Hyeon05PNAS}
Hyeon, C., and D.~Thirumalai, 2005.
\newblock Mechanical unfolding of {RNA} hairpins.
\newblock \emph{Proc. Natl. Acad. Sci. U. S. A.} 102:6789--6794.

\bibitem[Behrouzi et~al.(2012)Behrouzi, Roh, Kilburn, Briber, and
  Woodson]{Behrouzi12Cell}
Behrouzi, R., J.~H. Roh, J.~D. Kilburn, R.~M. Briber, and S.~A. Woodson, 2012.
\newblock Cooperative tertiary interaction network guides {RNA} folding.
\newblock \emph{Cell} 149:348--357.

\bibitem[Honeycutt and Thirumalai(1992)]{Honeycutt92Biopolymers}
Honeycutt, J.~D., and D.~Thirumalai, 1992.
\newblock The nature of folded states of globular proteins.
\newblock \emph{Biopolymers} 32:695--709.

\bibitem[Aronovitz and Nelson(1986)]{Aronovitz86JP}
Aronovitz, J.~A., and D.~R. Nelson, 1986.
\newblock Universal features of polymer shapes.
\newblock \emph{Journal de physique} 47:1445--1456.

\bibitem[Dima and Thirumalai(2004)]{Dima04JPCB}
Dima, R.~I., and D.~Thirumalai, 2004.
\newblock Asymmetry in the shapes of folded and denatured states of proteins.
\newblock \emph{J. Phys. Chem. B} 108:6564--6570.

\bibitem[Honeycutt and Thirumalai(1989)]{Honeycutt89JCP}
Honeycutt, J.~D., and D.~Thirumalai, 1989.
\newblock Static properties of polymer chains in porous media.
\newblock \emph{J. Chem. Phys.} 90:4542--19.

\bibitem[Soto et~al.(2007)Soto, Misra, and Draper]{Soto07Biochem}
Soto, A.~M., V.~Misra, and D.~E. Draper, 2007.
\newblock Tertiary structure of an {RNA} pseudoknot is stabilized by
  ``diffuse'' {M}g$^{2+}$ ions.
\newblock \emph{Biochemistry} 46:2973--2983.

\bibitem[Lee and Thirumalai(2001)]{Lee01Macromolecules}
Lee, N., and D.~Thirumalai, 2001.
\newblock Dynamics of collapse of flexible polyelectrolytes in poor solvents.
\newblock \emph{Macromolecules} 34:3446--3457.

\bibitem[Muthukumar(2004)]{Muthukumar04JCP}
Muthukumar, M., 2004.
\newblock Theory of counter-ion condensation on flexible polyelectrolytes:
  Adsorption mechanism.
\newblock \emph{J. Chem. Phys.} 120:9343--9350.

\bibitem[Chremos and Douglas(2016)]{Chremos16SoftMatter}
Chremos, A., and J.~F. Douglas, 2016.
\newblock Counter-ion distribution around flexible polyelectrolytes having
  different molecular architecture.
\newblock \emph{Soft Matter} 12:2932--2941.

\bibitem[Denesyuk et~al.(2018)Denesyuk, Hori, and Thirumalai]{Denesyuk18JPCB}
Denesyuk, N.~A., N.~Hori, and D.~Thirumalai, 2018.
\newblock Molecular simulations of ion effects on the thermodynamics of {RNA}
  folding.
\newblock \emph{J. Phys. Chem. B} 122:11860--11867.

\bibitem[Hayes et~al.(2015)Hayes, Noel, Mandic, Whitford, Sanbonmatsu, Mohanty,
  and Onuchic]{Hayes15PRL}
Hayes, R.~L., J.~K. Noel, A.~Mandic, P.~C. Whitford, K.~Y. Sanbonmatsu,
  U.~Mohanty, and J.~N. Onuchic, 2015.
\newblock Generalized {M}anning condensation model captures the {RNA} ion
  atmosphere.
\newblock \emph{Phys. Rev. Lett.} 114:258105--258106.

\end{thebibliography}


\newpage

\section*{Supplementary Material}



\includegraphics[width=0.8\textwidth]{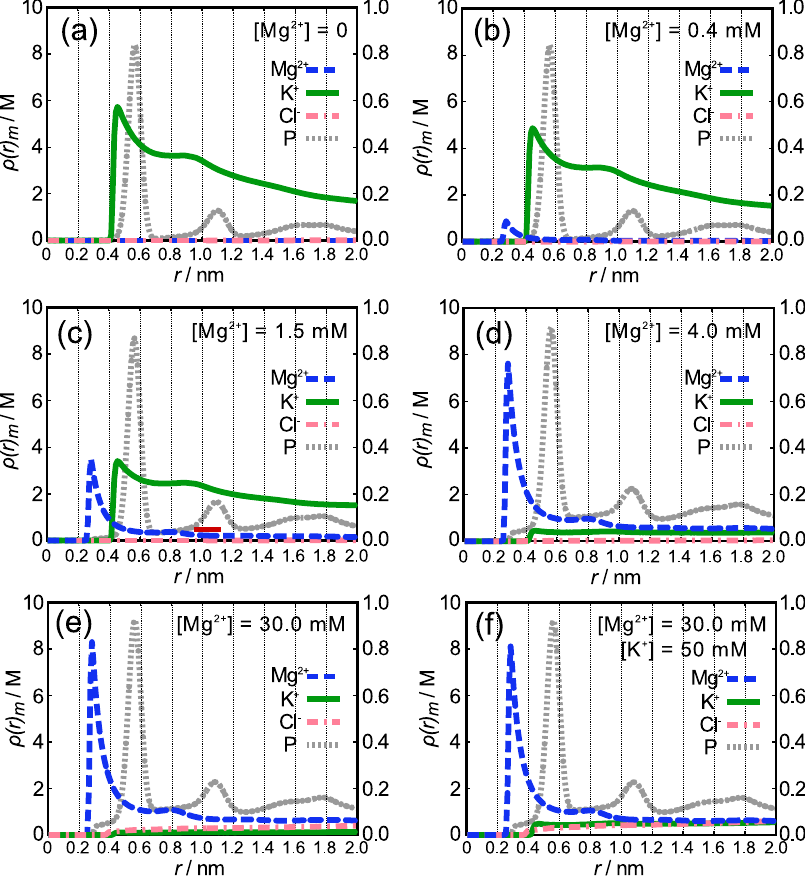}

\textbf{Figure S1:} Radial profiles of local ion concentration, $\rho(r)$,
for $\textrm{Mg}^{2+}$, $\textrm{K}^{+}$, $\textrm{Cl}^{-}$, and
phosphate (P) at {[}\K{}{]} = 12 mM and various {[}\Mg{}{]}: (a) 0, (b) 0.4 mM, (c) 1.5 mM, (d) 4.0 mM, and (e) 30 mM. 
(f) The profile at {[}\K{}{]} = 50 mM and {[}\Mg{}{]} $=$ 30 mM. 
Appropriate amount of $\textrm{Cl}^{-}$ were added to the solution for the electroneutrality.
Note that the scale for $\rho_{_{\text{Mg}^{2+}}}$ and $\rho_{_{\textrm{P}}}$ (left axis) is 10 times greater than $\rho_{_{\textrm{K}^{+}}}$ and $\rho_{_{\textrm{Cl}^{-}}}$ (right axis).

\clearpage{}

\includegraphics[width=0.8\textwidth]{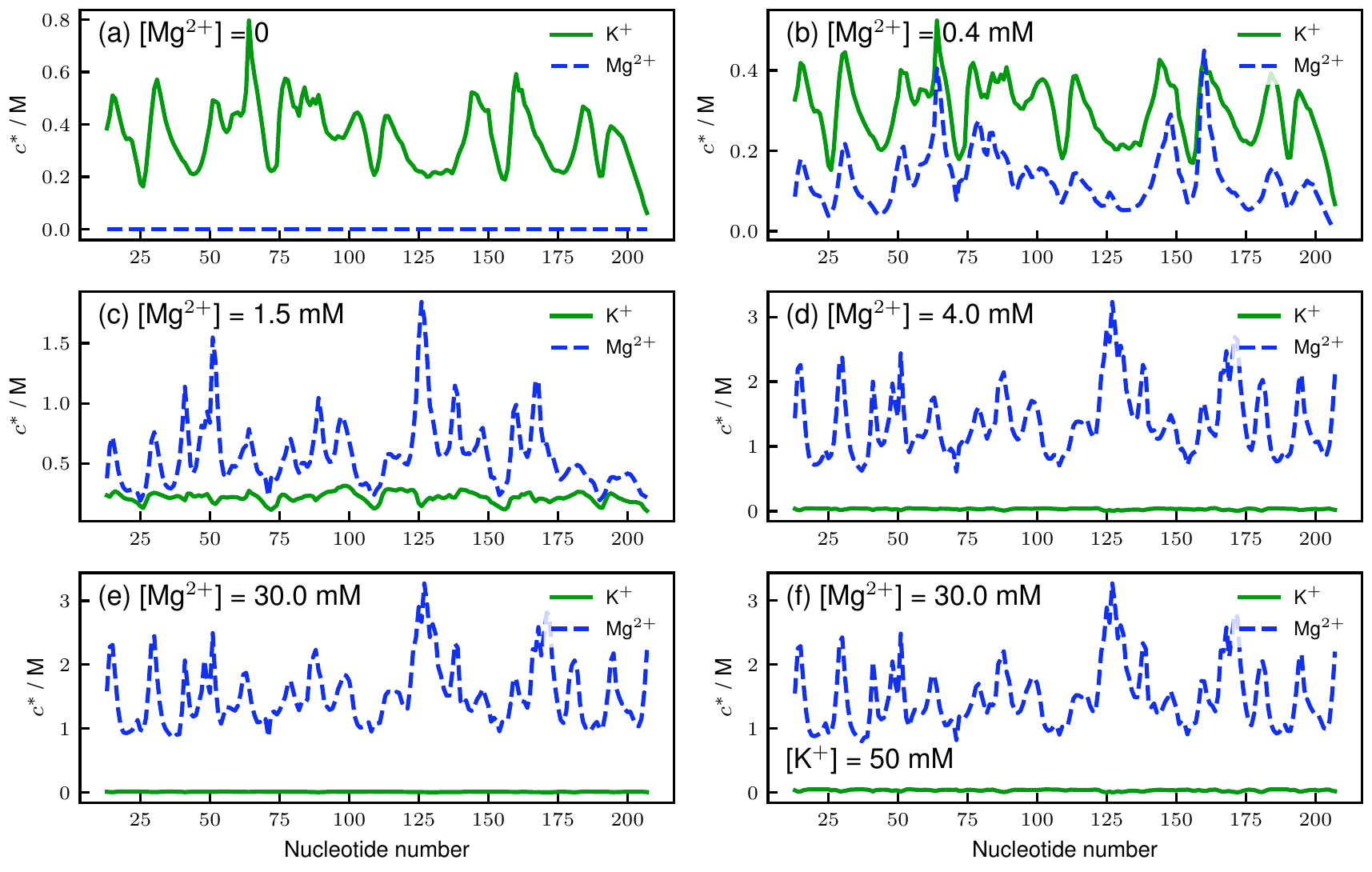}

\textbf{Figure S2:} Local ion concentrations, $c_{i}^{\ast}$ ($i$ is the
nucleotide position), for $\textrm{Mg}^{2+}$ (blue) and $\textrm{K}^{+}$
(green) at {[}\K{}{]} = 12 mM and various [\Mg{}]: (a) 0, (b) 0.4 mM, (c) 1.5 mM, (d) 4.0 mM, and (e) 30 mM.
(f) The profile at {[}\K{}{]} = 50 mM and [\Mg{}] $=$ 30 mM.

\clearpage{}

\includegraphics[width=0.9\textwidth]{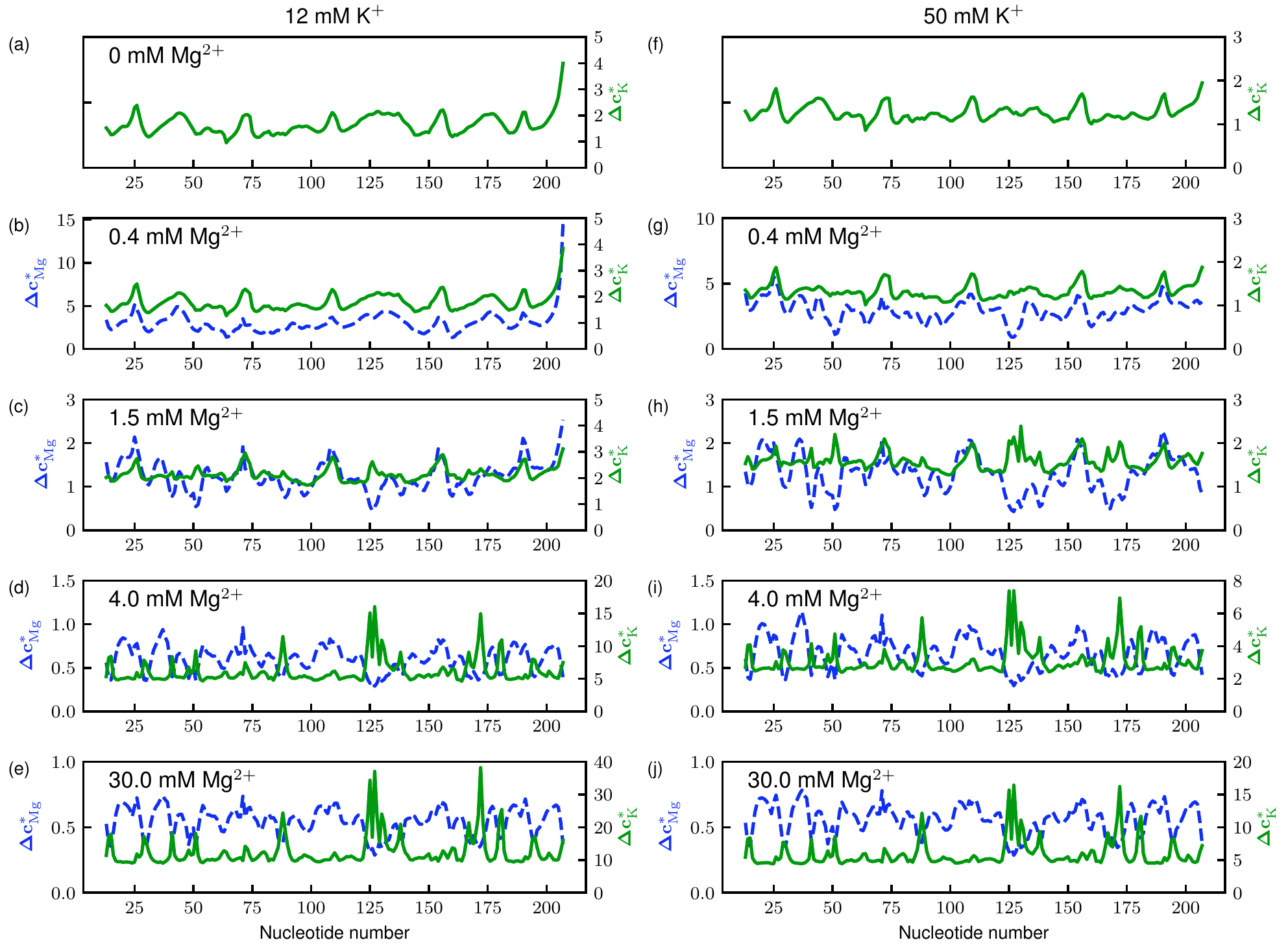}

\textbf{Figure S3:} Normalized fluctuations of local ion concentrations, $\Delta c^{\ast}$
(Eq.~4), for \Mg{} (dotted blue)
and \K{} (solid green), in 12 mM KCl (left column; a--e) and 50 mM KCl (right column; f--j).

\clearpage{}

\includegraphics[width=0.9\textwidth]{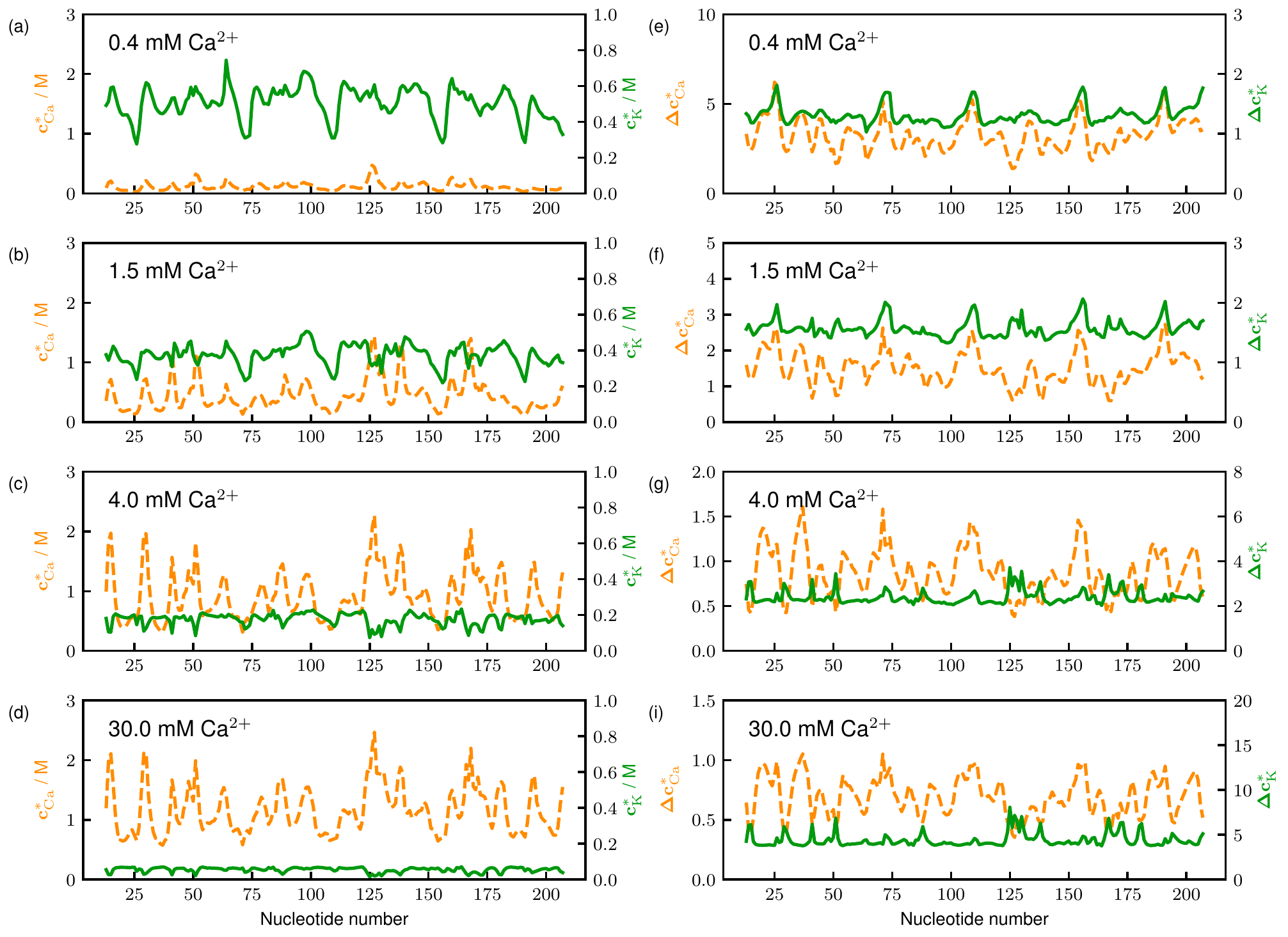}

\textbf{Figure S4:} Sequence dependence of the local ion concentration (left
column, $c^{\ast}$) and its fluctuations (right column, $\Delta c^{\ast}$)
in the presence of \Ca{} ions instead of \Mg{}. The concentration
of \Ca{} is labeled in each panel. {[}KCl{]} is fixed at 50 mM. 

\clearpage{}

\includegraphics[width=0.65\textwidth]{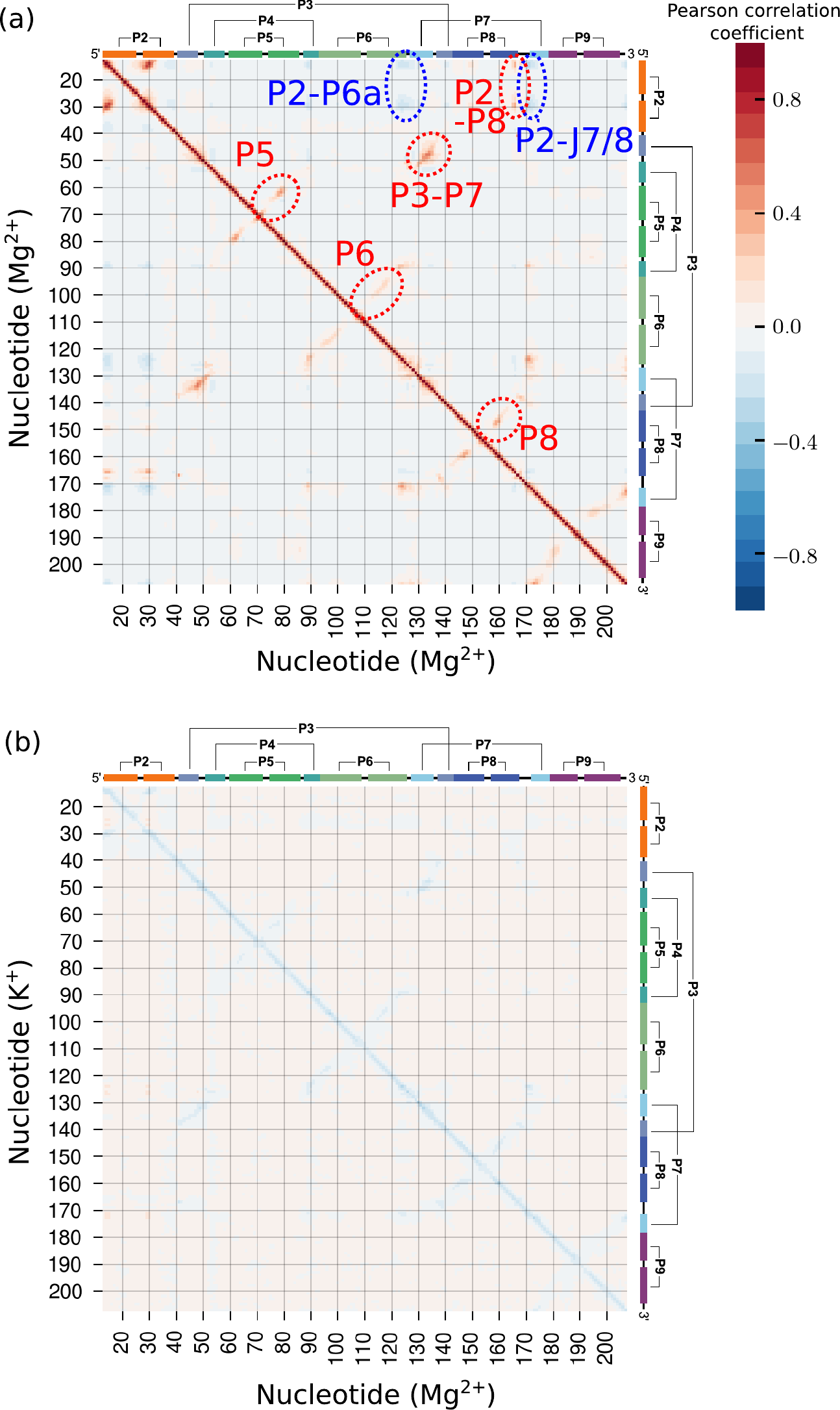}

\textbf{Figure S5:} \textbf{(a)} Correlations between \Mg{} condensations to two different nucleotides. The horizontal and vertical axes are the nucleotide positions $i$ and $j$. Color in each cell ($i$, $j$) represents the Pearson correlation coefficient (scale on the right) of the number of bound ions at site $i$ and $j$. The \Mg{} concentration is 1.5 mM and [\K{}] = 50~mM. \textbf{(b)} Correlations between \Mg{}-\K{} condensations is most prominent along the diagonal. This implies that \K{} ion is released upon \Mg{} binding.

\clearpage{}

\begin{table}[h!]

\textbf{Table S1: Parameters for excluded volume and electrostatic interactions.}
\\
\\
\centering
\begin{tabular}{lcccccccccccc}
Bead type &&&& $R_{i} / \textrm{nm} \,^\ast$ &&&& $\varepsilon_{i} / $kcal mol$^{-1}$ &&&& $z_{i} / e$\tabularnewline
\hline 
Phosphate &&&& 0.21 &&&& 0.2 &&&& -1\tabularnewline
Rod charge &&&& 0.21 &&&& 0.2 &&&& -1\tabularnewline
Sugar &&&& 0.29 &&&& 0.2 &&&& 0\tabularnewline
Base A &&&& 0.28 &&&& 0.2 &&&& 0\tabularnewline
Base G &&&& 0.30 &&&& 0.2 &&&& 0\tabularnewline
Base C &&&& 0.27 &&&& 0.2 &&&& 0\tabularnewline
Base U &&&& 0.27 &&&& 0.2 &&&& 0\tabularnewline
Mg$^{2+}$ &&&& 0.08 &&&& 0.9 &&&& 2\tabularnewline
Ca$^{2+}$ &&&& 0.17 &&&& 0.5 &&&& 2\tabularnewline
Cl$^{-}$ &&&& 0.19 &&&& 0.3 &&&& -1\tabularnewline
K$^{+}$ &&&& 0.27 &&&& 0.0003 &&&& 1\tabularnewline
\end{tabular}
\\
$^\ast$ If both interacting sites are RNA sites, we take $R_i+R_j=0.32$ nm.
\end{table}

\end{document}